\documentclass[aps,prb,reprint,superscriptaddress,longbibliography,floatfix]{revtex4-2}
\usepackage[utf8]{inputenc}
\usepackage[T1]{fontenc}
\usepackage{amsmath,amssymb,bm,mathtools}
\usepackage{graphicx}
\usepackage{physics}
\usepackage{booktabs,array,longtable,tabularx}
\usepackage[dvipsnames]{xcolor}
\usepackage{enumitem}
\usepackage{microtype}
\usepackage{circuitikz}
\setlist[itemize]{nosep,leftmargin=1.4em}
\setlist[enumerate]{nosep,leftmargin=1.6em}

\newcommand{\kap}{\kappa}
\newcommand{\nt}{n_t}
\newcommand{\nE}{n_E}
\newcommand{\nS}{n_S}
\newcommand{\nB}{n_B}
\newcommand{\Et}{\mathcal{E}_t}
\newcommand{\Ft}{\mathcal{F}_t}
\newcommand{\gEH}{g_E^{(H)}}
\newcommand{\gXH}{g_X^{(H)}}
\newcommand{\lt}{\ell_t}
\newcommand{\hwc}{\hbar\omega_c}
\newcommand{\gE}{g_E}
\newcommand{\gX}{g_X}
\newcommand{\avg}[1]{\langle #1 \rangle}
\newcommand{\lam}{\lambda}
\newcommand{\lamo}{\lambda_0}
\newcommand{\etaD}{\eta_D}
\newcommand{\dds}{\,\mathrm{d}s}

\begin{document}

% Replace the author and affiliation placeholders before submission.
\title{Thermodynamic Charge Partition in Accumulation-Layer Heterostructures}
\author{Elmar B\"ockenhoff}
\affiliation{Independent Researcher}
\email{eboeckenhoff@web.de}
\date{\today}

\begin{abstract}
We develop a thermodynamic description of accumulation-layer heterostructures in which the induced sheet density is partitioned between the near-interface accumulation-layer charge and a complementary screening charge in the surrounding structure. Treating this partition as the central state variable yields a complete Helmholtz free energy, a corrected locked-branch chemical potential, and a shifted release potential that separates energetic path selection from geometric capacitance. The physical path is selected spectrally: compressible segments remain fully screened, whereas incompressible segments evolve along a locked branch until release is triggered by the relevant gap. Differential capacitance, tunnel current and plateau width then emerge as different projections of the same coupled thermodynamic structure. A canonical two-stage self-consistent Poisson--Schr\"odinger reduction supplies universal master functions for the isolated accumulation layer and master surfaces for its finite-buffer extension, making the theory calculable across density and geometry. Comparison with magnetocapacitance and magnetotunneling data supports a picture in which nearby extended charge refills the accumulation layer and the effective screening depth grows with magnetic field.
\end{abstract}

\maketitle

% =====================
% Main text
% =====================

\section{Introduction}

Quantized accumulation layers in semiconductor heterostructures~\cite{AndoFowlerStern1982,Davies1998}
provide a setting in which thermodynamics, electrostatics, and quantum
structure are inseparably coupled. In capacitance and  tunneling measurements, the relevant response is not determined by the two-dimensional
electron system alone, but by the full heterostructure that stores, screens,
and redistributes the induced charge. This becomes especially important once
the accumulation-layer density of states develops a gap: charge can still be
induced externally even when continuous absorption into the accumulation layer
is no longer possible, and it is then no longer obvious where the additional
charge is stored or how the corresponding chemical-potential rise is
generated.

\begin{figure}
    \centering
    \includegraphics[width=0.9\linewidth]{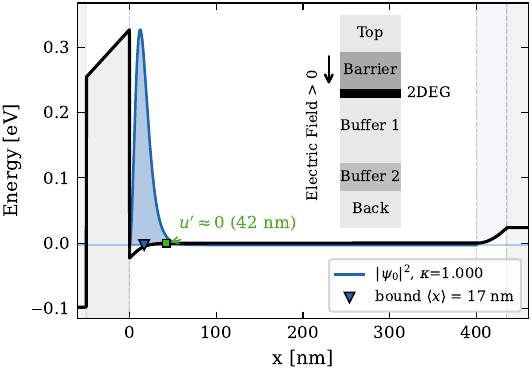}
    \caption{Conduction-band edge of the full heterostructure on the compressible branch ($\xi=0.8$, $\kappa_{\rm phys}=1.000$), corresponding to the conventional fully screened accumulation-layer picture. The blue fill shows the bound-state probability density $|\psi_0|^2$ at eigenenergy $\varepsilon_0$, with centroid $\langle x\rangle_{\rm bnd}=17\,\mathrm{nm}$ (blue $\blacktriangledown$). All induced charge resides in this bound state; no extended states are occupied. The green square marks the screening boundary, defined by $u'(s)<2\%$, at $x\simeq 42\,\mathrm{nm}$. The overlay indicates the layer sequence of the heterostructure and the positive electric-field direction.}
    \label{fig:band_compressible}
\end{figure}

A natural starting point is the density-based viewpoint introduced by
Hohenberg and Kohn~\cite{HohenbergKohn1964}, made operational by Kohn and
Sham~\cite{KohnSham1965} and extended to finite temperature by
Mermin~\cite{Mermin1965}, with formal foundations in
Refs.~\cite{Levy1979,Lieb1983}. Rather than prescribing an
\emph{a priori} potential profile, we formulate the state of the
heterostructure through its induced charge distribution, reduced here to the
physically relevant sheet-density variables
\begin{equation}
  n_t = n_E + n_S ,
  \label{eq:intro_partition}
\end{equation}
where \(n_t\) is the total induced sheet density, \(n_E\) the
accumulation-layer contribution, and \(n_S\) the complementary screening
contribution in the remainder of the heterostructure. This charge partition
is the central state variable of the present work; Fig.~\ref{fig:band_compressible}
shows the conduction-band profile for the compressible reference state
$\kappa = 1$, where the conventional fully screened picture applies.

\begin{figure}
    \centering
    \includegraphics[width=0.95\linewidth]{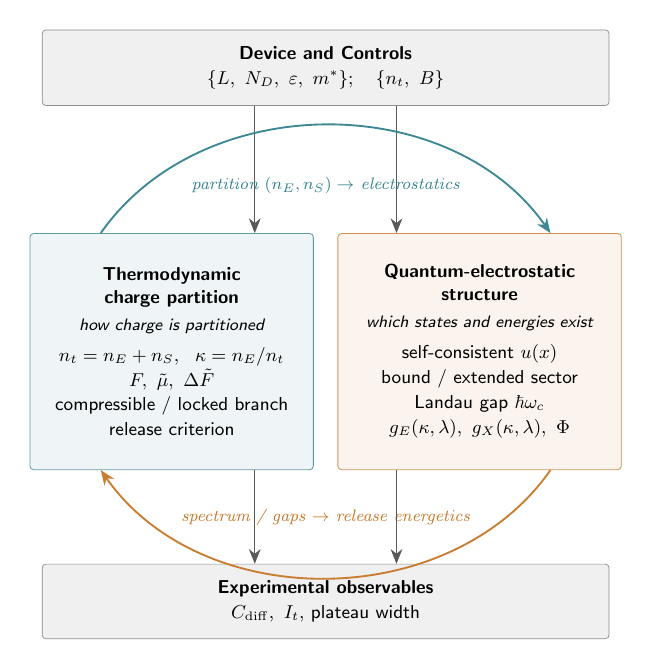}
    \caption{Schematic overview of the coupled thermodynamic and quantum-electrostatic structure of the problem. The external device and control parameters $(L,N_D,\varepsilon,m^\ast;n_t,B)$ define a heterostructure in which the total induced charge is partitioned thermodynamically into a bound accumulation-layer contribution and a remote screening contribution, $n_t=n_E+n_S$. This partition determines the electrostatic boundary conditions of the self-consistent Poisson--Schr\"odinger problem, which in turn provides the microscopic spectrum, the Landau gap, and the canonical response functions entering the release energetics. The central feedback loop therefore couples \emph{how charge is partitioned} to \emph{which states and energies exist}. Differential capacitance $C_{\rm diff}$, tunnel current $I_t$ and the incompressible plateau width all emerge as observables of this same coupled structure.}
\label{fig:thermo_ps_feedback}
\end{figure}

The heterostructure free energy is introduced in order to evaluate the
chemical potential of the \emph{entire} coupled structure. Once a spectral
gap prevents the accumulation layer from absorbing additional charge
continuously, the additional induced charge must be stored elsewhere in the
heterostructure. The associated incremental free-energy cost produces the
chemical-potential rise that must ultimately match the relevant spectral
threshold for release. In this way, the thermodynamic part of the problem
determines how charge is partitioned and stored, while the
quantum-electrostatic Poisson--Schr\"odinger part determines which states
and gaps exist. Differential capacitance, tunnel current and incompressible plateau width are then different projections
of this same coupled structure. Figure~\ref{fig:thermo_ps_feedback}
summarizes how the thermodynamic and Poisson--Schr\"odinger parts of the
problem feed into one another and into the observables.

The paper develops four interlocking contributions. (i) A Helmholtz
free-energy formulation of the complete heterostructure under the partition
Eq.~\eqref{eq:intro_partition}, with branch structure (compressible vs.\
locked) and a release condition $\Delta\mu=\Delta_{\rm gap}$ that follows
from the chemical-potential increment along the locked branch. (ii) A
canonical two-stage self-consistent Poisson--Schr\"odinger reduction in which
the isolated accumulation layer is represented by universal master functions
$g_E(\kappa)$, $g_X(\kappa)$, and the finite-buffer multisubband
heterostructure by canonical response surfaces on the $(\kappa,\lambda)$
grid; this canonical machinery underlies the entire framework and is
developed in detail in Appendix~\ref{sec:canonical_ps}. (iii) A prediction
of the differential capacitance from the thermodynamic theory, with
comparison against magnetocapacitance traces. (iv) A two-filter reduction
(spatial $\times$ spectral) of the perpendicular tunneling current, with
comparison against magnetotunneling data.

Section~\ref{sec:twodensity} formulates the heterostructure free energy and
the charge-partition variable. Section~\ref{sec:chemical_potential_from_F}
derives the chemical potential and identifies compressible and locked
branches. Section~\ref{sec:canonical_release_function} gives the canonical
release function and plateau width.
Sections~\ref{sec:three_energies} and~\ref{sec:tunnel_cond_experiment}
develop the differential capacitance and tunneling current, respectively,
with comparison to experiment. Section~\ref{sec:discussion} discusses scope
and connections, and Section~\ref{sec:conclusion} concludes.

\section{Thermodynamic description of the complete heterostructure}
\label{sec:twodensity}

\subsection{Why a free-energy description is needed}
\label{sec:why_free_energy}

A gapped accumulation layer poses a charge-redistribution problem for the
heterostructure as a whole. As long as the relevant accumulation-layer states
remain compressible, additional imposed charge can enter the mobile
two-dimensional sector directly. Once the Fermi level lies in a gap of the
in-plane spectrum, however, further charge can no longer be absorbed there
continuously. The heterostructure must still accommodate that charge, and the
corresponding incremental work must be stored elsewhere until the resulting
chemical-potential rise is large enough to overcome the gap.

In the present theory this problem is reduced to its minimal nontrivial form.
We distinguish a mobile accumulation-layer component with sheet density $n_E$
from a complementary screening component $n_S$ stored in the adjacent buffer.
The total imposed sheet density is therefore
\begin{equation}
  n_t = n_E + n_S,
  \qquad
  \kappa = \frac{n_E}{n_t},
  \qquad
  1-\kappa = \frac{n_S}{n_t}.
  \label{eq:nt_partition}
\end{equation}
Here $n_E$ carries the in-plane spectral structure of the accumulation layer,
while $n_S$ is the screening channel that remains available when the
accumulation-layer spectrum is gapped. The partition $\kappa$ is the state
variable of the charge-redistribution problem.

The role of the heterostructure free energy follows. It is needed to evaluate
the chemical potential of the \emph{entire} active structure and, in
particular, the incremental free-energy buildup associated with redirecting
charge away from the gapped accumulation layer. This stored increment produces
a chemical-potential rise $\Delta\mu$ that is later compared with the relevant
spectral threshold $\Delta_{\rm gap}$. The free energy is thus the
thermodynamic potential required to evaluate the release condition
$\Delta\mu=\Delta_{\rm gap}$.

It is useful to separate the total free energy into a passive stack part and
an active part,
\begin{equation}
  F_{\mathrm{tot}}
  =
  F_{\mathrm{stack}}(T,A,n_t)
  +
  F_{\mathrm{act}}(T,A,n_t,\kappa,B).
  \nonumber
\end{equation}
The first term contains those parts of the device that merely transmit the
imposed charge---the barrier and contact depletion regions---and are
insensitive to the partition. The second term contains the active subsystem:
the accumulation zone together with the compensating electrostatic channel of
the adjacent buffer. This active part carries the nontrivial
$\kappa$-dependence and therefore the storage energetics.

\begin{figure}
    \centering
    \includegraphics[width=0.95\linewidth]{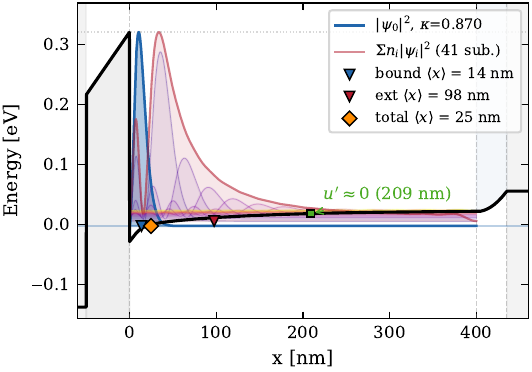}
    \caption{Conduction-band edge of the full heterostructure (barrier, buffer~1, buffer~2, back contact) on the locked branch at mid-plateau ($\xi=1.15$, $\kappa_{\rm phys}=0.870$). Blue fill: bound-state probability density $|\psi_0|^2$ at eigenenergy $\varepsilon_0$. Colored fills: extended-subband densities $n_i|\psi_i|^2$ at their respective eigenenergies (plasma colormap, 41 occupied subbands; magnified by a factor of 35 for visibility). Markers indicate the centroids of the bound sector, $\langle x\rangle_{\rm bnd}=14\,\mathrm{nm}$ (blue $\blacktriangledown$), the extended sector, $\langle x\rangle_{\rm ext}=98\,\mathrm{nm}$ (red $\blacktriangledown$), and the total charge distribution, $\langle x\rangle_{\rm tot}=25\,\mathrm{nm}$ (orange $\blacklozenge$). The green square marks the screening boundary, $u'(s)<0.02$ (with $u'(0)=1$ at the interface).}
    \label{fig:band_states}
\end{figure}

Equation~\eqref{eq:nt_partition} defines a two-dimensional state
space spanned by $(n_t,\kappa)$, but at fixed magnetic field the system traces
only a one-dimensional trajectory through it as $n_t$ is varied. We refer to
this realized trajectory as the \emph{physical path} and write it as
$\kappa=\kappa_{\mathrm{phys}}(n_t,B)$. The physical path is selected by the
in-plane spectrum rather than by an unconstrained minimization of $F$ over
$\kappa$ at fixed $n_t$. As long as the topmost occupied accumulation-layer
level is partially filled, additional imposed charge enters the bound sector
continuously and the path lies on the \emph{compressible branch} $\kappa=1$.
Once the Fermi level falls into a gap, the bound density can no longer grow
continuously; it is pinned at the integer-filling value, and surplus imposed
charge is redirected into the screening channel. The path is then forced onto
a \emph{locked branch} with $\kappa<1$, named for the locking of the mobile
density. The locked branch persists until the chemical-potential rise
accumulated along it equals the relevant spectral gap, at which point the
next level becomes accessible and the path rejoins a compressible branch. The
matching condition $\Delta\mu=\Delta_{\rm gap}$ that triggers this rejoining
is the \emph{release condition}. The free energy evaluated along the physical
path,
\begin{equation}
  F_{\mathrm{path}}(n_t)
  \equiv
  F_{\mathrm{tot}}\bigl(n_t,\kappa_{\mathrm{phys}}(n_t,B)\bigr),
  \label{eq:Fpath_def}
\end{equation}
with $T$, $A$, $B$ understood as parameters, then defines the thermodynamic
chemical potential
\begin{equation}
  \mu_{\mathrm{tot}}
  =
  \frac{1}{A}\frac{dF_{\mathrm{path}}}{dn_t}.
  \nonumber
\end{equation}
The free energy supplies the energetic landscape of the coupled structure;
the spectrum decides which compressible and locked segments compose the
physical path and where each terminates. Figure~\ref{fig:band_states} shows
the locked-branch counterpart to Fig.~\ref{fig:band_compressible}: the
bound interfacial state coexists with a Fermi-weighted extended screening
density distributed across the buffer.

This formulation does not provide an unconstrained equilibrium theory of the
partition at each $n_t$, nor a microscopic kinetics of branch switching,
hysteresis, or metastability. It provides a state function for the active
subsystem, the chemical-potential rise accumulated along the realized path,
and a controlled way of combining that stored work with the spectral
structure of the two-dimensional channel. From this combination the release
condition, the voltage partition, the differential capacitance, and the
tunneling response all follow.

\subsection{Device geometry: the complete heterostructure}
\label{sec:device_geometry_complete}

Self-consistent treatments of semiconductor inversion and accumulation
layers, going back to Stern's original work~\cite{Stern1972} and the
GaAs--GaAlAs heterojunction analysis of Stern and Das
Sarma~\cite{SternDasSarma1984}, established the framework on which the
present canonical formulation rests.
The heterostructure considered here corresponds to the experimental samples
analyzed in Sec.~\ref{sec:comparison_experiment}, retained at the level of
the three structural features that govern the charge partition. The
electrostatic stack is bounded by quasimetallic electrodes whose internal
charging is not addressed here. Between these boundaries the device consists
of three regions arranged from top to bottom:
\begin{enumerate}
  \item a \textbf{barrier} of thickness $d_b$,
  \item an active \textbf{buffer~1} of width $L$ containing the accumulation zone and, for completeness, a donor density $N_D$,
  \item and a compensating \textbf{buffer~2} of width $L^{+}$, donor density $N_D^{+}$, and permittivity $\varepsilon_A$, which restores charge neutrality of the slab of total width $L+L^{+}$.
\end{enumerate}
Sample-specific features beyond this level---modulation doping, spacer layers,
nonuniform barriers, or asymmetric contacts---enter only through smooth,
partition-independent contributions to the voltage drop. The canonical
reduction below separates these from the universal interfacial confinement
physics.

Material parameters such as $m^\ast$ and $\varepsilon$ enter through the
similarity scales
\begin{equation}
\begin{aligned}
  \Ft&=\frac{e^2}{\varepsilon}\nt,
  \qquad
  \lt=\left(\frac{\hbar^2}{2m^\ast\Ft}\right)^{1/3},
  \\
  \Et&=\Ft\lt,
  \qquad
  \lam=\frac{L}{\lt},
  \qquad
  \lam^{+}=\frac{L^{+}}{\lt},
  \end{aligned}
  \nonumber
\end{equation}
so that the structure of the self-consistent solution is encoded by
dimensionless canonical objects rather than by material-specific numbers.
The material parameters set the scales but do not enter the structural
statements.

The electrostatic free energy of the device decomposes as
\begin{equation}
  F_{\mathrm{tot}}
  =F_{\mathrm{bar}}+F_{\mathrm{buf1}}+F_{\mathrm{buf2}}.
  \label{eq:F_tot_stack}
\end{equation}
Only $F_{\mathrm{buf1}}$ carries the charge-partition physics; the barrier
and buffer~2 are needed for the measurable gate voltage and the device
capacitance baseline, but they are insensitive to the partition at fixed
$\nt$. This separation lets the active subsystem be treated microscopically
while preserving the full device context.

For later use we record the smooth noncanonical contributions per unit area.
For a barrier with geometric capacitance per unit area and the buffer~2
closure, one has
\begin{align}
  \frac{F_{\mathrm{bar}}}{A}
  &=\frac{(e\nt)^2}{2C_{\mathrm{geo}}}
   =\frac{e^2\nt^2d_b}{2\varepsilon_b},
  \label{eq:Fbar_classical}\\
  \frac{F_{\mathrm{buf2}}}{A}
  &=\frac{e^2\nt^3}{6\varepsilon_A N_D^{+}}.
  \label{eq:Fbuf2_classical}
\end{align}
These terms are smooth functions of $\nt$ and independent of $\kap$; all
field- and gap-induced physics resides in the active buffer~1 region.

In what follows the quasimetallic electrodes are treated only as external
boundary conditions, while the barrier and buffer~2 are carried along as
smooth stack elements outside the $\kap$-dependent canonical dynamics. Their
contribution is restored when the buffer-1 physics is converted into the
measurable gate voltage and device capacitance in
Sec.~\ref{sec:voltage_partition}.

\begin{figure}
    \centering
    \includegraphics[width=0.95\linewidth]{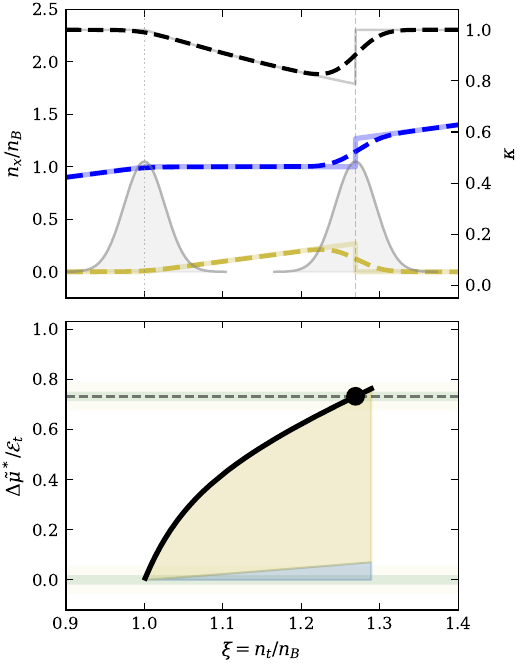}
    \caption{Partition mechanics across a representative plateau interval. Top panel: evolution of the charge partition and of the realized path $\kappa_{\rm phys}(\xi)$ as functions of the total induced density $\xi=n_t/n_D$. The dotted vertical lines mark the entry into the locked branch and the release point. The gray profiles indicate the neighboring Landau-level spectral weight. Bottom panel: accumulated dimensionless release contribution $\Delta\tilde{\mu}^{*}/\mathcal{E}_t$ along the locked branch. Starting from the compressible boundary, the release term grows monotonically, reaches the threshold marked by the dashed horizontal line, and thereby determines the endpoint of the plateau branch (black dot). The shaded area emphasizes the buildup of the release correction during the locked evolution.}
    \label{fig:partition_mechanics}
\end{figure}

\subsection{Complete Helmholtz free energy of the heterostructure}
\label{sec:complete_helmholtz}

The stack decomposition Eq.~\eqref{eq:F_tot_stack} places the entire
$\kap$-dependence in $F_{\mathrm{buf1}}(T,A,\nt,\kap;B)$, since only buffer~1
resolves how the imposed charge is partitioned between the bound and extended
sectors. Within the active subsystem there are three contributions. The first
is the vertical self-consistent confinement and electrostatic energy of
buffer~1, represented canonically by the stage-2 dimensionless free-energy
map $\widetilde F_{\mathrm{PS}}(\kap,\lam,\etaD)$. The second is the in-plane
free energy of the mobile accumulation-layer two-dimensional electron system
(2DES), which depends on $\nE=\kap\nt$ and on magnetic field. The third is
the in-plane free energy of the charge stored in the screening channel,
$\nS=(1-\kap)\nt$. The screening-channel states are quasi-three-dimensional;
their energetic separation is vanishingly small on the scale of the active
buffer, so Landau quantization is not resolved and no filling singularities
arise. We therefore write
\begin{equation}
\begin{aligned}
  F_{\mathrm{buf1}}
  &=A\nt\Et(\nt)\,\widetilde F_{\mathrm{PS}}(\kap,\lam,\etaD)\\
   &+A\,f_{\parallel}(\nE,B,T)
   +A\,f_{\parallel}^{(S)}(\nS,B,T).
  \end{aligned}
  \nonumber
\end{equation}
The canonical reduction stores all vertical buffer-1 physics in a
dimensionless master map; the in-plane spectra of the two electronic sectors
enter only through the surface-density functionals $f_{\parallel}$ and
$f_{\parallel}^{(S)}$. The symmetric roles of $\nE$ and $\nS$ on the in-plane
side mirror their symmetric appearance in the electrostatic sum
$\sum_i(n_i/\nt)\varepsilon_i$ entering $\widetilde F_{\mathrm{PS}}$, and
Sec.~\ref{sec:screening_channel_smallness} quantifies the magnitude of the
$f_{\parallel}^{(S)}$ contribution in the physically relevant regime. The
stage-1 and stage-2 self-consistent solutions of
Appendix~\ref{sec:canonical_ps} provide canonical realizations of these
partitioned states, indexed by $\kap$ and, in the finite-buffer problem,
also by $\lam$ and $\etaD=N_D\lt/\nt$.

At fixed $\nt$ the partition-dependent free-energy difference is
\begin{align}
  \Delta F(T,A,\nt,\kap;B)
  &\equiv F_{\mathrm{tot}}(\kap)-F_{\mathrm{tot}}(\kap=1)
  \nonumber
  \\
  &=A\nt\Et\,\bigl[\widetilde F_{\mathrm{PS}}(\kap,\lam,\etaD)-
      \widetilde F_{\mathrm{PS}}(1,\lam,\etaD)\bigr]
  \nonumber
  \\
  &\quad +A\bigl[f_{\parallel}(\kap\nt,B,T)-f_{\parallel}(\nt,B,T)\bigr]
  \nonumber
  \\
  &\quad +A\,f_{\parallel}^{(S)}\bigl((1-\kap)\nt,B,T\bigr). \nonumber
  \end{align}
The barrier and buffer~2 cancel from this difference because at fixed $\nt$
they do not depend on $\kap$. The screening-channel term $f_{\parallel}^{(S)}$
enters only on the locked branch ($\kap<1$), since $f_{\parallel}^{(S)}$
vanishes at $\nS=0$ on the compressible branch. The full structure must be
included to define the physical device, but the thermodynamic competition
between compressible and locked partitions is controlled entirely by the
active buffer~1 subsystem together with the in-plane free energies of the two
electronic sectors.

\subsection{The active-buffer free energy in canonical form}
\label{sec:active_buffer_free_energy}

The stage-2 canonical map provides the buffer contribution in the compact form
\begin{equation}
  \widetilde F_{\mathrm{PS}}(\kap,\lam,\etaD)
  = \sum_i \frac{n_i}{\nt}\,\varepsilon_i
    -\frac12\int_0^{\lam}\bigl[u'(s)\bigr]^2\,\dds.
  \label{eq:F_PS_canonical}
\end{equation}
The eigenvalue sum contains the self-consistent one-dimensional band energy of
the accumulated and extended subbands, while the integral subtracts the Hartree
double counting. The first term may be decomposed as
\begin{equation}
  \sum_i \frac{n_i}{\nt}\,\varepsilon_i
  =\kap\,\varepsilon_0 + E_{\mathrm{ext}},
  \qquad
  E_{\mathrm{ext}}=\sum_{i\ge 1}\frac{n_i}{\nt}\,\varepsilon_i.
  \nonumber
\end{equation}
The free-energy map therefore separates naturally into a bound contribution, an
extended-state contribution, and the Hartree correction. This decomposition is
the energetic counterpart of the microscopic density separation discussed later.

Equation~\eqref{eq:F_PS_canonical} compresses all $\kap$-dependent physics
of the heterostructure into a canonical subsystem that can be embedded into
the smooth classical stack. This is the level at which the thermodynamic
competition is decided.

\subsection{Microscopic picture of the charge partition}
\label{sec:micro_charge_partition}

\begin{figure}
    \centering
    \includegraphics[width=0.95\linewidth]{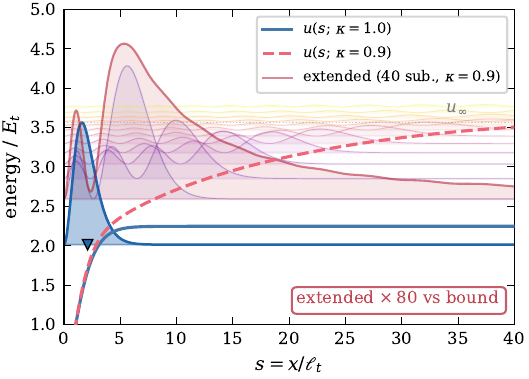}
    \caption{Stage-2 canonical solution at $\lambda_1=54$. The solid blue curve shows the band edge $u(s;\kappa=1.00,\lambda_1=54)$ on the fully screened branch, while the dashed red curve shows $u(s;\kappa=0.90,\lambda_1=54)$ after charge transfer into the screening channel. Wavefunctions are shown for $\kappa=0.90$ only: blue fill, bound-state probability density $|\psi_0|^2$; colored fills (plasma colormap), extended-subband densities $n_i|\psi_i|^2$; red line, total extended density envelope. The asymptotic shelf at $u_\infty$ is indicated. The text box indicates the visibility magnification of the extended contribution relative to the bound state.}
    \label{fig:stage2_canonical}
\end{figure}

The partition reflects a concrete reorganization of the self-consistent
buffer solution into two sectors with different spatial and energetic
character. The bound contribution $\nE$ is concentrated near the interface
and is carried predominantly by the lowest state $\psi_0$. The screening
contribution $\nS$ is built from higher subbands whose structure is
constrained by orthogonality to $\psi_0$; it does not form a second sharp
sheet at the interface, but appears as a broad, Fermi-weighted density
distributed over the wider accumulation zone and buffer.

The higher states are expelled from the interfacial core by orthogonality
and act as a nonlocal screening channel. Their incoherent sum generates the
extended density that carries $\nS$, while the near-interface region remains
dominated by the bound contribution $\nE$. The free-energy separation into
bound and extended terms therefore mirrors an actual decomposition of the
self-consistent wave mechanics, not an \emph{ad hoc} thermodynamic ansatz.
Figure~\ref{fig:stage2_canonical} contrasts the canonical solutions at
$\kappa = 1$ and $\kappa = 0.90$, making the bound-state localization and
the extended screening tail directly visible.

The observables of later sections---capacitance and tunneling---do not probe an abstract partition parameter. They probe this
self-consistent reorganization, in which a tightly bound interfacial sector
coexists with and exchanges charge with a broad extended sector. The
detailed self-consistent origin of the two sectors is developed in the
canonical framework, where the orthogonality minimum, the incoherent
Fermi-weighted density, and the screening decomposition are made explicit
in Sec.~\ref{sec:microscopic_anatomy}.

\section{Chemical potential from the free energy}
\label{sec:chemical_potential_from_F}

\begin{figure}
    \centering
    \includegraphics[width=0.95\linewidth]{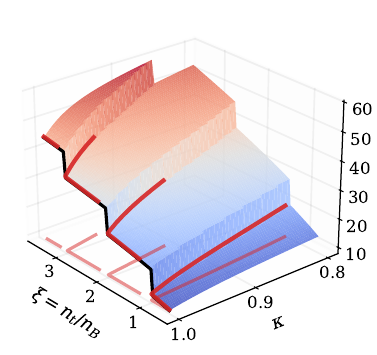}
    \caption{Chemical potential landscape $\mu(\xi,\kappa)$ in meV at $B=5.0$\,T, $L=200$\,nm, $\gamma=0$, and $T=0$, shown as a function of total induced density $\xi=n_t/n_D$ and partition parameter $\kappa$. The surface combines the $\hbar\omega_c$ staircase with the release correction $\Delta\mu^*$ and thereby visualizes the splitting into compressible and locked branches. Transparent gray planes indicate the Landau-level energies. The realized path $\kappa_{\rm phys}(\xi)$ therefore appears as a sequence of compressible segments connected by locked branches that terminate at the release points.}
    \label{fig:mu_xi_surface}
\end{figure}

The central thermodynamic quantity is the chemical work required to increase
the imposed charge along the realized path. Because the full heterostructure
free energy has already been written as a smooth stack contribution plus an
active $\kap$-dependent buffer contribution, the chemical potential can now
be derived in a form that separates generic device electrostatics from the
storage and release physics, yielding the chemical-potential rise of the
full coupled structure---the incremental quantity that must match the
spectral gap. Figure~\ref{fig:mu_xi_surface} shows the resulting
$\mu(\xi,\kappa)$ landscape, which combines the $\hbar\omega_c$ staircase
with the release correction $\Delta\mu^*$ and exposes the
compressible/locked branch structure of the realized path.

\subsection{General chain rule along a physical path}
\label{sec:mu_chain_rule}

At fixed area, the chemical potential is obtained by differentiating the
free energy along the realized path, not by varying the partition
independently at fixed $n_t$:
\begin{equation}
  \mu_{\mathrm{tot}}(\nt,B,T)
  = \frac{1}{A}\frac{\dd F_{\mathrm{tot}}}{\dd \nt}.
  \nonumber
\end{equation}
Using the decomposition of Sec.~\ref{sec:twodensity} we write
\begin{equation}
  \mu_{\mathrm{tot}}(\nt,B,T)
  = \mu_{\mathrm{stack}}(\nt)
  + \mu_{\mathrm{buf}}(\nt,\kap,\lam,\etaD;B,T),
  \nonumber
\end{equation}
where the smooth stack baseline is
\begin{equation}
  \mu_{\mathrm{stack}}(\nt)
  \equiv \frac{1}{A}\frac{\dd}{\dd \nt}
  \bigl[F_{\mathrm{bar}}+F_{\mathrm{buf2}}\bigr].
  \nonumber
\end{equation}
With the smooth electrostatic energies
Eqs.~\eqref{eq:Fbar_classical}--\eqref{eq:Fbuf2_classical}, this gives
\begin{equation}
  \mu_{\mathrm{stack}}
  =\frac{e^2\nt}{C_{\mathrm{geo}}}
   +\frac{e^2\nt^2}{2\varepsilon_{A}N_D^{+}}.
  \nonumber
\end{equation}
These terms are important for the measurable gate voltage but are smooth
functions of $\nt$ alone and therefore do not generate the plateau
structure.

The active buffer free energy per unit area is written as
\begin{equation}
\begin{aligned}
  \frac{F_{\mathrm{buf}}}{A}
  &= \nt\Et\,\widetilde F_{\mathrm{PS}}(\kap,\lam,\etaD)
  + f_{\parallel}(\nE,B,T)
  + f_{\parallel}^{(S)}(\nS,B,T),\\ \nE&=\kap\nt,\ \nS=(1-\kap)\nt.
  \label{eq:Fbuf_area}
\end{aligned}
\end{equation}
Since the canonical scales satisfy
\begin{equation}
  \Et \propto \nt^{2/3},
  \qquad
  \lam = \frac{L}{\lt} \propto \nt^{1/3},
  \qquad
  \etaD = \frac{N_D\lt}{\nt} \propto \nt^{-4/3},
  \nonumber
\end{equation}
all canonical arguments move when the imposed charge changes. Defining the
mobile-charge response
\begin{equation}
  \kappa_d \equiv \frac{\dd \nE}{\dd \nt},
  \nonumber
\end{equation}
one has
\begin{equation}
  \frac{\dd\kap}{\dd\nt}=\frac{\kappa_d-\kap}{\nt},
  \qquad
  \frac{\dd\lam}{\dd\nt}=\frac{\lam}{3\nt},
  \qquad
  \frac{\dd\etaD}{\dd\nt}=-\frac{4\etaD}{3\nt}.
  \nonumber
\end{equation}
The total derivative of Eq.~\eqref{eq:Fbuf_area} therefore yields
\begin{align}
  \mu_{\mathrm{buf}}
  &= \frac{1}{A}\frac{\dd F_{\mathrm{buf}}}{\dd \nt}
  \nonumber \\
  &= \Et\biggl[
      \tfrac53\,\widetilde F_{\mathrm{PS}}
      +(\kappa_d-\kap)\pdv{\widetilde F_{\mathrm{PS}}}{\kap}
      +\tfrac{\lam}{3}\pdv{\widetilde F_{\mathrm{PS}}}{\lam}
      -\tfrac{4\etaD}{3}\pdv{\widetilde F_{\mathrm{PS}}}{\etaD}
    \biggr]
  \nonumber \\
  &\quad + \kappa_d\,\mu_{\parallel}^{(E)}(\nE,B,T)
         + (1-\kappa_d)\,\mu_{\mathrm{ext}}^{(S)}(\nS,B,T),
  \label{eq:mu_total_polished}
\end{align}
where
\begin{equation}
  \mu_{\parallel}^{(E)}(\nE,B,T)
  = \pdv{f_{\parallel}}{\nE},
  \qquad
  \mu_{\mathrm{ext}}^{(S)}(\nS,B,T)
  = \pdv{f_{\parallel}^{(S)}}{\nS}
  \nonumber
\end{equation}
are the in-plane chemical potential of the mobile two-dimensional
accumulation-layer sector and the chemical potential of the
quasi-three-dimensional screening channel, respectively. The weights
$\kappa_d$ and $1-\kappa_d$ follow from $\dd\nE/\dd\nt=\kappa_d$ and
$\dd\nS/\dd\nt=1-\kappa_d$. Equation~\eqref{eq:mu_total_polished} is the
central chain-rule statement. It separates five distinct contributions to
the chemical work: overall density scaling of the canonical energy, motion
through partition space $(\kap)$, motion through geometry $(\lam)$, motion
through the doping parameter $(\etaD)$, and the in-plane response of the
two electronic sectors. The branch analysis below identifies which of
these contributions survives on the compressible branch, which survive on
the locked branch, and how the stored increment builds the release
chemical potential.

\subsection{Compressible and locked branches}
\label{sec:compressible_locked_mu}

The physical path is assembled from two response laws selected by the
accumulation-layer spectrum. The compressible/incompressible distinction
goes back to the electrostatic edge-channel theory of Chklovskii \emph{et
al.}~\cite{Chklovskii1992} and to the self-consistent Hartree screening
calculations of a 2DEG in quantizing fields by Wulf, Gudmundsson, and
Gerhardts~\cite{WulfGudmundssonGerhardts1988,GudmundssonGerhardts1987}.
As long as the topmost Landau level can still
accept charge, the imposed charge continues to enter the mobile sector. The
system then remains on the compressible branch,
\begin{equation}
  \kappa_d = 1,
  \qquad \kap = 1,
  \qquad \frac{\dd\kap}{\dd\nt}=0,
  \nonumber
\end{equation}
so that Eq.~\eqref{eq:mu_total_polished} reduces to
\begin{equation}
\begin{aligned}
\mu_{\mathrm{buf}}^{\mathrm{comp}}
  = &\Et\left[
      \frac53\widetilde F_{\mathrm{PS}}
      + \frac{\lam}{3}\partial_{\lam}\widetilde F_{\mathrm{PS}}
      - \frac{4\etaD}{3}\partial_{\etaD}\widetilde F_{\mathrm{PS}}
    \right]_{\kap=1}\\
 & + \mu_{\parallel}^{(E)}(\nt,B,T).
  \end{aligned}
  \nonumber
\end{equation}
On this branch the accumulation layer itself absorbs the additional charge,
so no storage term associated with transfer into the screening channel is
yet present.

Once the Fermi level enters a gap of the in-plane spectrum, the accumulation
layer can no longer absorb further charge continuously. The path is then
diverted onto a locked branch on which the mobile density is pinned while
the total imposed charge continues to change. In the sharp idealization this
gives
\begin{equation}
  \kappa_d = 0,
  \qquad
  \frac{\dd\kap}{\dd\nt}=-\frac{\kap}{\nt},
  \nonumber
\end{equation}
and the active-buffer contribution becomes
\begin{equation}
\begin{aligned} \mu_{\mathrm{buf}}^{\mathrm{lock}}
  = &\Et\left[
      \frac53\widetilde F_{\mathrm{PS}}
      - \kap\partial_{\kap}\widetilde F_{\mathrm{PS}}
      + \frac{\lam}{3}\partial_{\lam}\widetilde F_{\mathrm{PS}}
      - \frac{4\etaD}{3}\partial_{\etaD}\widetilde F_{\mathrm{PS}}
    \right]\\
  &+ \mu_{\mathrm{ext}}^{(S)}(\nS,B,T).
  \label{eq:mu_lock}
\end{aligned}
\end{equation}
On the locked branch additional imposed charge is redirected into the
screening channel, and the term
\begin{equation}
  -\kap\partial_{\kap}\widetilde F_{\mathrm{PS}}
  \nonumber
\end{equation}
measures the corresponding locked-branch release contribution. It appears
because the mobile population is frozen while the total imposed charge and
its associated canonical scales continue to evolve. The system can therefore
keep storing imposed charge even though the mobile accumulation-layer
density no longer changes.

The full device chemical potential is obtained by adding back the smooth
stack term,
$\mu_{\mathrm{tot}}^{\mathrm{comp/lock}}=\mu_{\mathrm{stack}}+
\mu_{\mathrm{buf}}^{\mathrm{comp/lock}}$. Since the stack contribution is
smooth and branch-independent, the branch competition is decided entirely
by the active subsystem. The spectrum therefore first determines whether
the system remains compressible or enters the locked branch, and the locked
branch then accumulates the chemical work that is later tested against the
spectral release threshold (Fig.~\ref{fig:physical_path}).

\subsection{Quasi-three-dimensional screening channel: magnitude of the correction}
\label{sec:screening_channel_smallness}

The in-plane contribution $f_{\parallel}^{(S)}$ of the screening channel in
Eq.~\eqref{eq:mu_lock} is vanishingly small. We show here that, at the
physical device parameters, this term corrects the locked-interval widths
$\delta_\nu$ by less than $0.1\%$ and can therefore be dropped from the
release condition. A perturbative reinstatement is given at the end of the
subsection.

Because the screening-channel electrons occupy a macroscopic region of the
heterostructure (thickness $W_S$ of order
$10^2$--$10^3\,\mu\mathrm{m}$), their single-particle spectrum is a dense
quasi-continuum on the scale of the Landau gap:
$\Delta E_S = \hbar^2\pi^2/(2m^\ast W_S^{\,2}) \sim 10^{-9}\,\hwc$ for GaAs at
$B=5\,\mathrm{T}$. Landau-level structure is therefore not resolved and the
in-plane free energy takes the Sommerfeld form
\begin{equation}
  f_{\parallel}^{(S)}(\nS) = \frac{\nS^{\,2}}{2D_S},
  \qquad
  \mu_{\mathrm{ext}}^{(S)}(\nS) = \frac{\nS}{D_S},
  \label{eq:fS_quadratic}
\end{equation}
with $D_S=D_{3\mathrm{D}}W_S$ the effective surface density of states.
Equation~\eqref{eq:fS_quadratic} is independent of $B$ and $T$ at leading
order. Its coupling to the release condition is controlled by the single
dimensionless number
\begin{equation}
  \gamma_S \equiv \frac{\nB}{D_S\,\hwc},
  \nonumber
\end{equation}
the ratio of the Landau degeneracy $\nB=eB/h$ to the number of
screening-channel states per Landau energy. On the sharp locked branch
[$\kap=\nu/\xi$, $\nS=(\xi-\nu)\,n_D$, $\xi\equiv \nt/n_D$, with $n_D$ the
degeneracy per resolved level; see Sec.~\ref{sec:degeneracy_gap_role}] the
release criterion of Sec.~\ref{sec:canonical_release_function} becomes
\begin{equation}
  d_F(\kap,\lam,\etaD)\,\alpha_C\,\xi^{2/3} + \gamma_S(\xi-\nu) = 1,
  \label{eq:release_with_screening_channel}
\end{equation}
where $\alpha_C=C_{\mathrm{E}}\nB^{2/3}/\hwc$. At $\gamma_S=0$ this reduces
to the canonical release equation used in the rest of the paper.

For GaAs at $B=5\,\mathrm{T}$ we have $\nB\approx
1.2\times10^{15}\,\mathrm{m^{-2}}$ and $\hwc\approx 8.6\,\mathrm{meV}$. At
the representative locked point $n_S\approx 0.27\,\nB$ the three-dimensional
density is $n_{3\mathrm{D}}=n_S/W_S\sim 10^{17}{-}10^{18}\,\mathrm{m^{-3}}$,
and using $D_{3\mathrm{D}}\propto n_{3\mathrm{D}}^{1/3}$ one obtains
\begin{equation}
  \gamma_S \approx 7\times 10^{-4}\text{--}2\times 10^{-3}
  \quad\text{for}\quad W_S\in[300,1000]\,\mu\mathrm{m}.
  \label{eq:gammaS_estimate}
\end{equation}

Numerical solution of Eq.~\eqref{eq:release_with_screening_channel} for the
reference device of Sec.~\ref{sec:canonical_release_function} confirms
three features over the range $\gamma_S\in[0,0.2]$, which extends two
orders of magnitude beyond the physical value of
Eq.~\eqref{eq:gammaS_estimate}. First, the effect is small even at the
tabulated maximum: the relative change of $\delta_\nu$ between
$\gamma_S=0$ and $\gamma_S=0.2$ ranges from $-8.7\%$ at $\nu=1$ to $-3.7\%$
at $\nu=5$, decreasing monotonically with $\nu$. Linear extrapolation to
the physical value $\gamma_S\sim 10^{-3}$ gives a correction of order
$5\times 10^{-4}$ on $\delta_1$, i.e.\ below $0.1\%$---well below the
disorder- and map-related uncertainties and unable to modify the plateau
structure at any level relevant to experiment. Second, the effect is
perturbative across the entire range: the numerical slope
$-\partial\delta_\nu/\partial\gamma_S$ near $\gamma_S=0$ takes the values
$0.130$ at $\nu=1$ and $0.070$ at $\nu=2$, consistent with the
linearization
$\delta_\nu(\gamma_S)\approx
  \delta_\nu(0)[1-\gamma_S/d_F'(\xi_{\mathrm{rel}}^{(0)})]$,
and changes by less than $10\%$ out to $\gamma_S=0.2$. Third, $\delta_\nu$
is monotonically decreasing in $\gamma_S$ for every $\nu$ and approaches
the expected strong-coupling asymptote $\delta_\nu\to 1/\gamma_S$, at which
the plateaus collapse---the correct physical outcome when the screening
channel is infinitely soft.

Accordingly, we set $f_{\parallel}^{(S)}=0$ for the remainder of the paper.
The correction can be reinstated at any stage by the replacement
\begin{equation}
  d_F\,\alpha_C\,\xi^{2/3}\;\longrightarrow\;
  d_F\,\alpha_C\,\xi^{2/3}+\gamma_S(\xi-\nu)
  \nonumber
\end{equation}
in the release condition, which is linear and additive in $\gamma_S$.

\subsection{The one-sided entry offset and the corrected release potential}
\label{sec:shifted_release_potential}

The quantity in Eq.~\eqref{eq:mu_lock} is still not the final plateau
storage cost. The locked branch is entered through a one-sided structural
offset at $\kap\to 1^-$. Exactly at $\kap=1$ the system is a single-component
state: only the bound accumulation state is occupied, and the extended sector
is empty. For $\kap\to 1^-$, even an infinitesimal extended occupation opens a
new screening channel and changes the self-consistent solution everywhere in
buffer~1. The chemical potential therefore has a one-sided limit
\begin{equation}
  \mu(1^-,\lam,\etaD),
  \nonumber
\end{equation}
which measures the entry cost of establishing the two-component branch.

The physically relevant plateau-storage quantity is the additional chemical
work accumulated \emph{after} this entry cost has been paid. We therefore
define the shifted release potential
\begin{equation}
  \Delta\mu^*(\kap,\lam,\etaD)
  = \mu_{\mathrm{buf}}^{\mathrm{lock}}(\kap,\lam,\etaD)
    - \mu_{\mathrm{buf}}^{\mathrm{lock}}(1^-,\lam,\etaD).
  \label{eq:dmu_star}
\end{equation}
By construction, $\Delta\mu^*$ vanishes at branch entry and rises as the
extended sector develops. It is therefore the stored chemical-potential
increment that must later be compared with $\Delta_{\rm gap}$. In the numerical
map this is precisely the role of the $\kap=1$ regularization: the locked-branch
limit is retained, whereas the true compressible value is stored separately and
used only on the compressible branch. This guarantees that the release function
starts from zero at onset.

\subsection{Density-space convolution smoothing}
\label{sec:smoothing}

In the sharp (zero-temperature, zero-disorder) limit, the electric
subband density~$n_E^{(0)}(n_t)$ is a staircase with discontinuities at
Landau-level edges~$n_t = \nu n_B$ and at the handover
intersections~$n_t = n_i(\nu)$. To account for the combined effect of
Landau-level broadening and thermal smearing, we convolve the sharp
staircase with a Gaussian kernel in density space:
\begin{equation}
  n_E(n_t) = \int_{-\infty}^{\infty}
    n_E^{(0)}(n_t')\,
    \frac{1}{\sqrt{2\pi}\,\sigma_n}
    \exp\!\Bigl(-\frac{(n_t - n_t')^2}{2\sigma_n^2}\Bigr)
    \,\mathrm{d}n_t',
  \nonumber
\end{equation}
with the screening density following from particle conservation,
$n_S(n_t) = n_t - n_E(n_t)$.

The density-space width $\sigma_n$ is derived from the total energy
broadening $\sigma_E$ by noting that each Landau level accommodates
$n_B = g_d eB/h$ electrons over an energy range $\hbar\omega_c$, where
$g_d$ is the spin/valley degeneracy:
\begin{equation}
  \sigma_n = \frac{\sigma_E}{\hbar\omega_c}\,n_B
           = \sigma_E \times \frac{g_d\, m^*}{2\pi\hbar^2}.
  \nonumber
\end{equation}
Here $\sigma_E$ combines disorder-induced Landau-level broadening $\Gamma$
and thermal smearing in quadrature:
\begin{equation}
  \sigma_E = \sqrt{\Gamma^2
    + \Bigl(\frac{\pi}{\sqrt{3}}\,k_{\mathrm{B}} T\Bigr)^{\!2}},
  \qquad
  \Gamma(B) = \gamma_{\mathrm{dis}}\sqrt{B},
  \nonumber
\end{equation}
where $\gamma_{\mathrm{dis}}$ is the disorder broadening coefficient
(in $\mathrm{meV/\sqrt{T}}$) and the factor $\pi/\sqrt{3}$ converts the
Fermi--Dirac thermal width to an equivalent Gaussian standard deviation.

\begin{figure}
    \centering
    \includegraphics[width=0.95\linewidth]{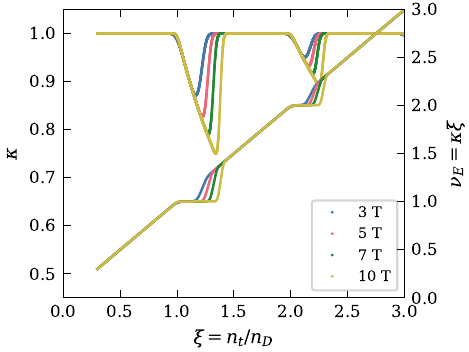}
    \caption{Realized physical path of the charge partition. Top set of
    curves (left axis): $\kappa_{\rm phys}$ as a function of the canonical
    filling parameter $\xi = n_t/n_D$ for $B \in \{3,5,7,10\}\,\mathrm{T}$.
    Compressible segments appear as $\kappa=1$; locked branches appear as
    $\kappa<1$ excursions in which the accumulation-layer density is pinned
    at $n_E=\nu n_D$ and the additional induced charge is absorbed by the
    screening channel. Bottom set of curves (right axis): the corresponding
    accumulation-layer filling $\nu_E = \kappa\xi$, which exhibits integer
    plateaus at $\nu = 1, 2$ and linear refilling between them. Increasing
    $B$ broadens and deepens the locked excursions because the Landau gap
    that must be overcome to trigger release grows with field.}
    \label{fig:physical_path}
\end{figure}

This single convolution smooths \emph{both} types of staircase
discontinuities---at the Landau-level edges and at the handover
intersections---without requiring separate treatment. The resulting smooth
functions $n_E(n_t)$ and $n_S(n_t)$ (see Figs.~\ref{fig:partition_mechanics}
and~\ref{fig:physical_path})
define the occupation fraction
$\kappa(n_t)$ [Eq.~\eqref{eq:nt_partition}], which enters the subband
energy $E_0(n_t, \kappa)$, the chemical potential $\mu_\parallel(n_E)$,
and all derived quantities (capacitance, tunneling current).

\section{The canonical release function and plateau width}
\label{sec:canonical_release_function}

Once the locked branch has been identified, the remaining question is how much
chemical work can be stored before release. When combined with the density
dependence of the canonical scales, the corrected locked-branch chemical
potential yields a fully canonical release function that expresses the
plateau width in terms of the same self-consistent map used for the
equilibrium analysis. The magnetic field and the origin of the gap then
enter only through a small set of dimensionless parameters.

\subsection{Locked-branch kinematics and the onset parameters}
\label{sec:locked_branch_kinematics}

At filling factor $\nu$, the locked branch begins when the topmost resolved
level has just become full. Throughout that locked interval the mobile density
is pinned at
\begin{equation}
  \nE = \nu n_D = \mathrm{const},
  \nonumber
\end{equation}
where $n_D$ is the degeneracy per resolved level. Writing
\begin{equation}
  \xi = \frac{\nt}{n_D},
  \qquad
  \kap = \frac{\nE}{\nt} = \frac{\nu}{\xi},
  \nonumber
\end{equation}
shows that the physical path through state space is fixed once the onset
condition $\xi=\nu$ is known. Let $\lt^{(\nu)}=\lt(\nu n_D)$ and
$\Et^{(\nu)}=\Et(\nu n_D)$ be the canonical scales at onset. Since
\begin{equation}
\begin{aligned}
  \lt(\xi) &= \lt^{(\nu)}\Bigl(\frac{\xi}{\nu}\Bigr)^{-1/3}
            = \lt^{(\nu)}\kap^{1/3},
  \\
  \Et(\xi) &= \Et^{(\nu)}\Bigl(\frac{\xi}{\nu}\Bigr)^{2/3}
            = \Et^{(\nu)}\kap^{-2/3},
  \label{eq:path_scales}
  \end{aligned}
\end{equation}
the dimensionless width of buffer~1 evolves as
\begin{equation}
  \lam(\kap)
  = \frac{L}{\lt(\xi)}
  = \frac{L}{\lt^{(\nu)}}\kap^{-1/3}
  = \lamo\kap^{-1/3},
  \label{eq:lambda_path}
\end{equation}
where
\begin{equation}
  \lamo \equiv \frac{L}{\lt^{(\nu)}}
  \nonumber
\end{equation}
is the onset width. Once the spectral condition has selected the locked
branch, the subsequent path is no longer free: it is fixed kinematically by
$\kappa=\nu/\xi$ together with the density dependence of the canonical
scales. The plateau problem therefore reduces to following a known
trajectory until the release condition is met.
Figure~\ref{fig:physical_path} shows the resulting $\kappa_{\rm phys}(\xi)$
across the first few plateaus for several fields, exhibiting both the
$\kappa=\nu/\xi$ kinematic shape of each locked branch and the
field-dependent excursion width.

\subsection{Release criterion in physical and canonical form}
\label{sec:release_criterion}

\begin{figure}
    \centering
    \includegraphics[width=0.95\linewidth]{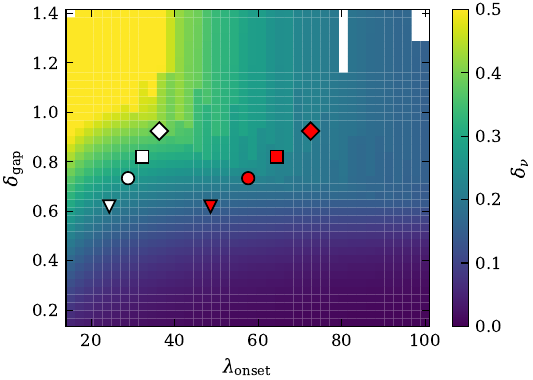}
    \caption{Universal plateau-width map showing the normalized plateau width $\delta_\nu$ as a function of the onset parameter $\lambda_{\rm onset}$ and the normalized gap $\delta_{\rm gap}=\hbar\omega_c/\mathcal{E}_t$. The color scale (viridis) spans $\delta_\nu=0$ to $0.5$. Symbols mark device operating points at $B=3$\,T ($\blacktriangledown$), $5$\,T ($\bullet$), $7$\,T ($\blacksquare$), and $10$\,T ($\blacklozenge$); white fill denotes $L=200$\,nm and red fill denotes $L=400$\,nm. The overlaid symbols locate the operating points of the devices discussed in the text within this universal parameter map.}
    \label{fig:plateau_width_map}
\end{figure}

Plateau release occurs when the accumulated locked-branch chemical work reaches
the relevant spectral threshold $\Delta_{\rm gap}$:
\begin{equation}
  \Delta\mu^*(\kap,\lam,\etaD)\,\Et(\xi)
  = \Delta_{\rm gap}.
  \nonumber
\end{equation}
Substituting Eqs.~\eqref{eq:path_scales} and \eqref{eq:lambda_path}
gives
\begin{equation}
  \Delta\mu^*\!\bigl(\kap,\lamo\kap^{-1/3},\etaD(\kap)\bigr)
  \,\Et^{(\nu)}\kap^{-2/3}
  = \Delta_{\rm gap}.
  \nonumber
\end{equation}
It is therefore natural to define the canonical release function
\begin{equation}
  \Phi(\kap,\lamo;\etaD)
  \equiv
  \Delta\mu^*\!\bigl(\kap,\lamo\kap^{-1/3},\etaD(\kap)\bigr)
  \kap^{-2/3}.
  \nonumber
\end{equation}
The release criterion becomes
\begin{equation}
  \Phi(\kap^*,\lamo;\etaD) = \delta_{\rm gap},
  \qquad
  \delta_{\rm gap} \equiv \frac{\Delta_{\rm gap}}{\Et^{(\nu)}},
  \label{eq:Phi_release_eq}
\end{equation}
where the asterisk denotes the release-point value: $\kap^*$ is the value of
$\kap$ at which the criterion is satisfied, and the corresponding density
parameter is $\xi^*=\nu/\kap^*$. The same convention applies to the shifted
release potential $\Delta\mu^*$ of Sec.~\ref{sec:shifted_release_potential}:
the asterisk marks quantities specific to the release condition. The
left-hand side of Eq.~\eqref{eq:Phi_release_eq} is determined entirely by
the canonical Poisson--Schr\"odinger map once the locked-branch trajectory
has been fixed; the right-hand side is a dimensionless measure of the
in-plane spectral gap. The evolution of the physical path is then fully
explicit: the spectrum first selects whether the system remains compressible
or enters the locked branch, and, once locked, the same spectrum provides
the threshold that terminates that branch.

Once $\kap^*$ is known, the plateau width in filling-factor units is
\begin{equation}
  \delta_\nu
  = \xi^* - \nu
  = \nu\left(\frac{1}{\kap^*}-1\right).
  \nonumber
\end{equation}
This expresses the plateau width directly in terms of the canonical release
function and the dimensionless gap parameter. Figure~\ref{fig:plateau_width_map}
shows the resulting universal plateau-width map
$\delta_\nu(\lambda_{\rm onset},\delta_{\rm gap})$, with the operating
points of the devices analyzed in Secs.~\ref{sec:three_energies}
and~\ref{sec:tunnel_cond_experiment} overlaid.

\subsection{What enters through the gap and degeneracy}
\label{sec:degeneracy_gap_role}

The orbital degeneracy per Landau level is $n_B=eB/h$, but the effective
degeneracy per resolved level can be $n_D=n_B$, $2n_B$, or another multiple
depending on whether spin, valley, or other internal degrees of freedom are
resolved. In the canonical release function the microscopic nature of the gap
enters only through two onset parameters: the geometric onset width
$\lamo=L/\lt(\nu n_D)$ and the dimensionless gap
$\delta_{\rm gap}=\Delta_{\rm gap}/\Et^{(\nu)}$. This separation is a key
structural simplification. The vertical self-consistent confinement problem
and the in-plane spectral gap remain distinct; Landau quantization provides
one important realization, but the canonical release formalism itself only
requires an in-plane gap.

\subsection{Reliability range and the map-edge regularization}
\label{sec:Phi_structure_reliability}

The release function satisfies
\begin{equation}
  \Phi(1^-,\lamo;\etaD)=0,
  \nonumber
\end{equation}
then rises as the extended sector fills, and typically reaches a maximum
before dropping again as $\kap$ becomes too small. This defines a natural
reliability range, with two regimes observed in the numerical map. In
Regime~I, $\kap\in[0.8,1]$, the canonical observables are smooth functions
of $1-\kap$ and the plateau problem is well behaved. In Regime~II, below
roughly $\kap\approx 0.8$, the discrete filling of higher subbands
introduces sharp features and the simple release-function picture becomes
less robust. The application of interest here lies within
Regime~I.

At $\kap=1$ itself the self-consistent map has a one-sided discontinuity
because the extended sector is exactly empty on the compressible slice but
nonzero on the locked side. The spline representation must therefore be
regularized by replacing the value at $\kap=1$ in the locked-branch map
with a linear extrapolation from the neighboring locked points. The true
compressible value is stored separately and used only for the compressible
branch. This step implements the subtraction of Eq.~\eqref{eq:dmu_star} at
the map level.

\section{Voltage partition and differential capacitance}
\label{sec:three_energies}

\subsection{Energetic path and geometric response}
\label{sec:energetic_vs_geometric}

The chemical-potential formula defines the energetic content of the path
through state space: how much chemical work is stored as the system moves
from the compressible branch onto the locked branch and toward release. The
measured capacitance, by contrast, is a geometric response around that
path. It depends on where the incremental charge sits, how the voltage is
partitioned across the heterostructure, and how nearby extended charge can
be exchanged with the accumulation layer. The two are related but distinct,
and the analysis that follows keeps them separate.

\subsection{Voltage partition across the full heterostructure}
\label{sec:voltage_partition}

The canonical buffer-1 solution developed above captures the entire
$\kap$-dependent physics relevant for the free energy, the chemical
potential, and the release condition. The remaining smooth parts of the device
--- the barrier and buffer~2 --- do not modify this intrinsic
$\kap$-dependent structure. Their role is instead to convert the intrinsic
buffer-1 response into the measurable gate voltage and to set the capacitance
baseline. The quasimetallic electrodes merely define the terminal voltage and
need not be discussed as thermodynamically active regions.

Accordingly, the total gate voltage may be decomposed as
\begin{equation}
\label{eq:Vgate-full}
  eV_g=
  \underbrace{\frac{e^2\nt}{C_{\rm geo}}}_{V_{\rm bar}}
  +\underbrace{\Et(\nt)\,\Delta V_1(\kap,\lam_1)}_{V_{1}}
  +\underbrace{\frac{e^2\nt^2}{2\varepsilon_A N_D^{+}}}_{V_{2}}.
\end{equation}
Here $V_{\rm bar}$ is the barrier drop, $V_{1}$ is the voltage drop across
the active buffer~1, and $V_{2}$ is the smooth depletion contribution of
buffer~2; $\Delta V_1$ denotes the full canonical voltage drop across
buffer~1.

The intrinsic thermodynamic quantities $\Delta F$, $\Delta\mu$,
$\Delta\mu^*$, and the plateau width are determined entirely by the active
buffer-1 problem and are therefore independent of the external closure
details. The measured gate voltage and the differential capacitance, by
contrast, depend on the full series combination of barrier, buffer~1, and
buffer~2 contributions in Eq.~\eqref{eq:Vgate-full}. The buffer-2 term is
smooth and $\kap$-independent; it may therefore be omitted during the
analysis of the intrinsic buffer-1 physics and restored only in the final
conversion to measurable quantities.

In differential form the same decomposition yields
\begin{equation}
  \frac{\dd V_g}{\dd \nt}
  =
  \frac{1}{e}\frac{\dd}{\dd \nt}\!\left(\frac{e^2\nt}{C_{\rm geo}}\right)
  +\frac{1}{e}\frac{\dd}{\dd \nt}\!\left[\Et(\nt)\Delta V_1(\kap,\lam_1)\right]
  +\frac{\dd V_{2}}{\dd \nt},
  \nonumber
\end{equation}
so that the differential capacitance may be written as
\begin{equation}
  C_{\rm diff}^{-1}
  =\frac{1}{e}\frac{\dd V_g}{\dd \nt}
  =C_{\rm bar}^{-1}+C_{\rm buf1}^{-1}+C_{\rm buf2}^{-1},
  \nonumber
\end{equation}
with $C_{\rm bar}=C_{\rm geo}$,
$C_{\rm buf1}^{-1}=(1/e)\,\dd[\Et\,\Delta V_1]/\dd\nt$, and
$C_{\rm buf2}^{-1}=\dd V_2/\dd(e\nt)$. The magnetic-field dependence is
concentrated in $C_{\rm buf1}^{-1}$, while the barrier and buffer~2
contributions supply a smooth baseline. The smooth closure can therefore
be ignored during the derivation of the intrinsic canonical release physics
and reintroduced only when laboratory observables are reconstructed.

\begin{figure}
    \centering
    \includegraphics[width=0.95\linewidth]{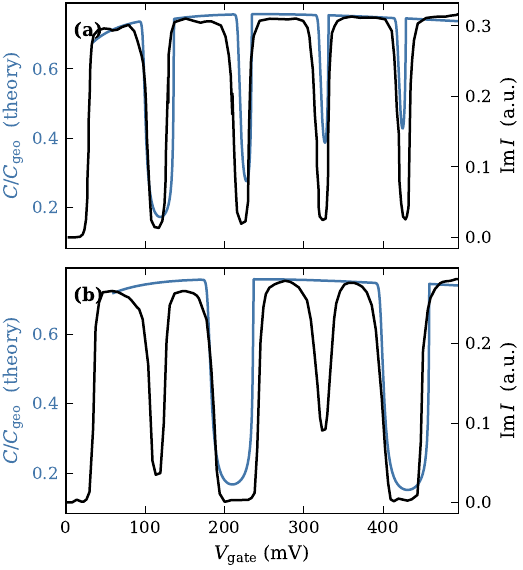}
    \caption{Comparison of calculated and measured capacitance traces as a function of gate voltage for the nominal buffer length $L=1000$\,nm. \textbf{(a)}~$B=2$\,T. \textbf{(b)}~$B=4$\,T. Blue curves (left axis): normalized capacitance $C/C_{\rm geo}$ from the present calculation for $\gamma=0.10$ and $T=0$. Black curves (right axis): experimental Im$\,I$ signal from scan 3\_10. A common multiplicative factor of $0.95$ was applied to the theoretical gate-voltage scale to account phenomenologically for the voltage drop across buffer~2 not captured by the nominal-geometry calculation. The nominal-geometry calculation captures the overall evolution of the capacitance trace, while quantitative deviations remain, especially in the detailed gap widths and in the high-field fine structure.}
    \label{fig:CvsVg_twopanel_B2T_B4T}
\end{figure}

\subsection{Differential centroid from the self-consistent map}
\label{sec:strict_differential_centroid}

Let $\avg{x}_{\mathrm{tot}}(\nt) \equiv \int_0^{L_1}\! x\,\rho(x)\,\dd x\,/\,\nt$
be the first moment of the buffer-1 charge density, so that the voltage drop
across buffer~1 is $V_1 = e\,\nt\,\avg{x}_{\mathrm{tot}}(\nt)/\varepsilon_A$.
The exact differential form of $1/C_{\mathrm{buf}1}$ is then
\begin{equation}
  \frac{1}{C_{\mathrm{buf}1}}
    = \frac{\dd V_1}{\dd(e\nt)}
    = \frac{1}{\varepsilon_A}\,
          \frac{\dd}{\dd \nt}\!\bigl[\nt\,\avg{x}_{\mathrm{tot}}(\nt)\bigr].
  \nonumber
\end{equation}
In canonical units, $\avg{x}_{\mathrm{tot}} = \lt(\nt)\,\avg{s}_{\mathrm{tot}}(\kap,\lam_1)$,
where $\avg{s}_{\mathrm{tot}}$ is the dimensionless first moment computed from
the self-consistent $u$-profile. Because the stage-2 self-consistent solver
operates at neutral boundary conditions $u'(0)=1$, $u'(\lam_1)=0$,
integration by parts gives the compact identity
\begin{equation}
  \avg{s}_{\mathrm{tot}}(\kap,\lam_1)
    = \int_0^{\lam_1}\! s\,\rho(s)\,\dd s
    = \int_0^{\lam_1}\! u'(s)\,\dd s,
  \nonumber
\end{equation}
so $\avg{s}_{\mathrm{tot}}$ is directly obtainable from the stored
$u'$-profile without further self-consistent iteration.

Along the physical path, $\lt$, $\kap$, and $\lam_1$ are all functions of
$\nt$. The scaling $\lt\propto \nt^{-1/3}$ gives
$\nt\,\dd\lt/\dd\nt = -\lt/3$. Defining the mobile-charge response
$\kappa_d \equiv \dd n_E/\dd \nt$, so that $n_E = \kap\,\nt$ implies
$\dd \kap/\dd\nt = (\kappa_d-\kap)/\nt$, and noting
$\dd \lam_1/\dd\nt = \lam_1/(3\nt)$, one obtains after straightforward
differentiation
\begin{equation}
\begin{aligned}
  \frac{\dd}{\dd \nt}\!\bigl[\nt\,\avg{x}_{\mathrm{tot}}\bigr]
    &= \lt\Bigl[\,
         \tfrac{2}{3}\,\avg{s}_{\mathrm{tot}}
         + (\kappa_d - \kap)\,\partial_{\kap}\avg{s}_{\mathrm{tot}} \\
    &\quad {}+ \tfrac{\lam_1}{3}\,\partial_{\lam_1}\avg{s}_{\mathrm{tot}}
       \Bigr].
\end{aligned}
  \label{eq:xincr_strict_general}
\end{equation}
Equation~\eqref{eq:xincr_strict_general} defines the \emph{strict}
incremental centroid $\avg{x}_{\mathrm{incr}}^{\mathrm{strict}}$ that enters
$C_{\mathrm{buf}1}^{-1} = \avg{x}_{\mathrm{incr}}^{\mathrm{strict}}/\varepsilon_A$.
It is valid on both branches and at their smooth crossover, with no
separate bound/extended decomposition.

On the compressible branch, $\kap=\kappa_d=1$ gives $(\kappa_d-\kap)=0$ and
$\avg{s}_{\mathrm{tot}}(1,\lam_1)=s_x(1,\lam_1)\approx\gX(1)$. The
$\partial_{\lam_1}\avg{s}_{\mathrm{tot}}$ term is numerically small because
the ground-state centroid is nearly $\lam_1$-independent, and one recovers
\begin{equation}
  \avg{x}_{\mathrm{incr}}^{\mathrm{strict,\,comp}}
    \approx \tfrac{2}{3}\,\lt\,\gX(1).
  \nonumber
\end{equation}
The factor $\tfrac{2}{3}$ is the classic triangular-well correction: the
strict differential differs from the electrostatic centroid by the
$\lt\propto\nt^{-1/3}$ response of the self-consistent length itself.

On the locked branch, $\kappa_d=0$ and $\kap = \nu/\xi$ is held
(approximately) fixed on the $\nu$-th plateau, so
Eq.~\eqref{eq:xincr_strict_general} reduces to
\begin{equation}
  \avg{x}_{\mathrm{incr}}^{\mathrm{strict,\,lock}}
    = \lt\!\left[\,
         \tfrac{2}{3}\,\avg{s}_{\mathrm{tot}}
         - \kap\,\partial_{\kap}\avg{s}_{\mathrm{tot}}
         + \tfrac{\lam_1}{3}\,\partial_{\lam_1}\avg{s}_{\mathrm{tot}}
       \right].
  \nonumber
\end{equation}
The term $-\kap\,\partial_{\kap}\avg{s}_{\mathrm{tot}}$ captures a
contribution that is invisible to the simpler blended ansatz
$\kappa_d\,\avg{x}_{\rm bound}+(1-\kappa_d)\,\avg{x}_{S}$: along the locked
branch, adding $\dd\nt$ at fixed $n_E$ forces $\kap$ to drop, and the
\emph{already-present} charge redistributes to maintain self-consistency at
the new $\kap$. This reorganization of the existing configuration shows up
in $\dd[\nt\avg{x}_{\mathrm{tot}}]/\dd\nt$ even though it is not a
new-electron placement.

The sharp $(\kappa_d,\kap)$ transition at plateau boundaries is broadened in
practice by the same disorder mechanism discussed in
Sec.~\ref{sec:smoothing}. In the strict formulation this broadening is
implemented by replacing the sharp path with a smoothed $n_E(\nt)$ and
computing $\kappa_d = \dd n_E/\dd\nt$ from it directly. Because $\kappa_d$
and $\kap$ already parameterize the branch structure, no additional
smoothing prescription is needed.

\subsection{Canonical capacitance statement and quantum capacitance}
\label{sec:canonical_capacitance_statement}

The capacitance can therefore be stated canonically: the energetic part of
the plateau problem is controlled by the release function $\Phi$, whereas
the differential response is controlled by the $\kap$-dependent loss of
bound-state screening and the resulting shift of the incremental charge
centroid. The two statements refer to the same canonical buffer solution,
but to different projections of it.

In the present formulation the usual quantum capacitance is not introduced
as an independent circuit element. Rather, it appears as a limiting form
of the thermodynamic capacitance obtained from the curvature of the
Helmholtz free energy. The thermodynamic origin of capacitance through
the electronic compressibility was introduced as a device concept by
Luryi~\cite{Luryi1988}; capacitance and magnetocapacitance measurements
have since been used to probe the 2DEG density of states
directly~\cite{Smith1985,Smith1986} and to reveal interaction effects
including negative compressibility~\cite{Eisenstein1994}, with more recent
extensions to multi-layer settings~\cite{Berthod2021}.
For a single electronic sheet density $n$, one may
write
\begin{equation}
  \frac{C_q}{A}
  = e^2\frac{\partial n}{\partial \mu}
  = e^2\left(\frac{\partial \mu}{\partial n}\right)^{-1}.
  \nonumber
\end{equation}
In the present heterostructure, however, the relevant state is described
by the partition Eq.~\eqref{eq:nt_partition}, and the measured differential
response is obtained from the reduced Helmholtz free energy along the
physical path, $F_{\rm path}(n_t)$ [Eq.~\eqref{eq:Fpath_def}].
Consequently,
\begin{equation}
  C_{\rm diff}^{-1}
  \propto
  \frac{\dd}{\dd n_t}\!\left[\frac{1}{A}\frac{\dd F_{\rm path}}{\dd n_t}\right].
  \nonumber
\end{equation}
This expression already contains the electronic compressibility, vertical
self-consistent confinement, geometric electrostatics, and charge transfer
between the accumulation layer and the screening channel. Adding a
separate quantum-capacitance element would therefore double-count part of
the same thermodynamic curvature. The usual $C_q$ is recovered only in
the limiting case where the accumulation-layer density is the sole active
charge coordinate. Figure~\ref{fig:CvsVg_twopanel_B2T_B4T} compares the
resulting $C_{\rm diff}(V_g)$ trace with magnetocapacitance data at $B=2$\,T
and $B=4$\,T.

\section{Tunneling current}
\label{sec:tunnel_cond_experiment}

\subsection{Tunneling as a spatial and spectral filter}
\label{sec:tunneling_from_accumulation_layer}

Magnetotunneling from accumulation layers~\cite{PhysRevB.38.10120,Boeckenhoff1988}
and from related heterostructure configurations~\cite{PhysRevB.39.1097,PhysRevB.40.8363}
has been used as a spectroscopic probe of subband and Landau-level structure.
The tunnel current probes the same charge partition that controls the
capacitance, but it probes it through a more selective operation. The
partition Eq.~\eqref{eq:nt_partition} separates the heterostructure charge
into the near-barrier accumulation-layer density $n_E$ and the more remote
screening density $n_S$. The tunneling process acts first as a
\emph{spatial filter}. Since the accumulation-layer charge is closest to
the barrier, its bare tunneling matrix element is exponentially larger
than that of the spatially displaced screening charge. If $d_E$ and $d_S$
denote the effective distances of the two charge sectors from the
barrier, with $d_S>d_E$, then
\begin{equation}
    \frac{|t_S|^2}{|t_E|^2}
    \sim
    \exp[-2q(d_S-d_E)]
    \ll 1 ,
  \nonumber
\end{equation}
where
\begin{equation}
    q=
    \frac{\sqrt{2m^\ast(V_0-E)}}{\hbar}
  \nonumber
\end{equation}
is the evanescent decay constant. Thus the tunnel experiment does not measure
the full density \(n_t\) directly. To leading order it filters out the
non-accumulation-layer charge and selects the \(n_E\) sector.

The same process also acts as a \emph{spectral filter}. Removing an electron
from the accumulation layer is a many-body operation: the remaining
\(N-1\)-electron state must be physically available on the physical
path. Thus the current is not only proportional to the amount of
near-barrier charge, but also to the energy-resolved many-body spectral
availability of the state left behind after tunneling. The screening charge
\(n_S\), although spatially filtered out of the bare tunneling amplitude,
therefore re-enters spectrally by changing the many-body configuration and by
shifting the available final-state spectral weight.

In this sense the tunnel current applies two filters in series:
\begin{equation}
    \text{tunneling}
    =
    \text{spatial filter selecting } n_E
    \quad
    \times
    \quad
    \text{many-body spectral filter along the physical path}.
  \nonumber
\end{equation}

We retain the simplifying assumption used throughout the tunneling simulation:
the bare tunneling coefficient is independent of the applied voltage over the
relevant bias interval. The magnetic-field dependence of the current is
therefore attributed not to a voltage-dependent barrier transparency, but to
the evolution of \(n_E(B)\) and to the many-body spectral availability of the
accumulation-layer states along the physical path.

The emitter is the lowest, and only occupied, accumulation-layer subband. In a
magnetic field perpendicular to the layer and parallel to the tunneling
direction, its in-plane density of states is split into Landau levels,
\begin{equation}
    E_{0j}
    =
    \varepsilon_0(n_t,\kappa)
    +
    \hbar\omega_c\left(j+\frac{1}{2}\right),
    \qquad
    j=0,1,2,\ldots ,
  \nonumber
\end{equation}
where \(\varepsilon_0\) is the self-consistent subband energy. Because the
subband is macroscopically occupied, the relevant energy is not the
single-particle subband eigenvalue alone, but the many-particle contribution
\begin{equation}
    E_{\rm sub}^{(0)}
    =
    N_E\varepsilon_0(n_t,\kappa),
    \qquad
    N_E=A n_E .
  \nonumber
\end{equation}
A tunneling event therefore removes an electron from an interacting
quasi-two-dimensional many-body state.

\subsection{Energy-resolved spectral availability}
\label{sec:tunnel_spectral_availability}

Let \(\hat c_{0j}\) remove an electron from Landau level \(j\) of the
accumulation-layer subband. The many-body matrix element associated with a
candidate final state is
\begin{equation}
    Z_{i\to f}^{(j)}
    =
    \left|
    \left\langle
        \Psi_f^{N-1}
        \middle|
        \hat c_{0j}
        \middle|
        \Psi_i^N
    \right\rangle
    \right|^2 ,
  \nonumber
\end{equation}
where \(|\Psi_i^N\rangle\) is the initial \(N\)-electron state and
\(|\Psi_f^{N-1}\rangle\) is the final state left behind after tunneling.
This matrix element is the many-body part of the spectral filter. It decides
whether the final state can be produced at all by removing an electron from
the accumulation-layer charge selected by the spatial filter.

The corresponding energy-resolved removal spectral function is
\begin{equation}
    A_E^{(j)}(\omega)
    =
    \sum_f
    Z_{i\to f}^{(j)}
    \delta\!\left[
        \omega-\left(E_i^N-E_f^{N-1}\right)
    \right].
  \nonumber
\end{equation}
The factor \(Z_{i\to f}^{(j)}\) gives the weight of an admissible many-body
final state, while the delta function places that weight at its proper energy.
In a finite sample the delta function is broadened by disorder, lifetime
effects, inelastic relaxation, or coupling to reservoirs,
\begin{equation}
    \delta(\omega-\omega_f)
    \rightarrow
    L_\gamma(\omega-\omega_f).
  \nonumber
\end{equation}
Thus the spectral filter is energy-resolved: an allowed state contributes
strongly only if its spectral weight lies near the tunneling energy.

With a voltage-independent bare tunneling prefactor, the current can be written
as
\begin{equation}
    I_t
    =
    I_\ast
    \sum_j
    n_j\,\mathcal W_j ,
    \label{eq:It_spectral_general}
\end{equation}
where \(n_j\) is the areal occupation of Landau level \(j\), \(I_\ast\)
contains the smooth barrier and collector factors, and
\begin{equation}
    \mathcal W_j
    =
    \sum_f
    Z_{i\to f}^{(j)}
    L_\gamma(\omega_{\rm tun}-\omega_f)
    \label{eq:Wj_microscopic}
\end{equation}
is the effective many-body spectral availability of level \(j\) at the
tunneling energy \(\omega_{\rm tun}\).

Equation~\eqref{eq:It_spectral_general} displays the two filters explicitly.
The factor \(I_\ast n_j\) represents the spatially selected occupation of the
accumulation-layer emitter, while \(\mathcal W_j\) represents the many-body
spectral filter imposed by the physical path.

For the numerical model, the microscopic object in
Eq.~\eqref{eq:Wj_microscopic} is reduced to a single effective displacement
scale \(\Delta_j\). We write
\begin{equation}
    \mathcal W_j
    \longrightarrow
    \mathcal W_j^{(\alpha)}
    =
    \mathcal W_{0,j}
    T_\alpha(\Delta_j),
  \nonumber
\end{equation}
with
\begin{equation}
    T_\alpha(\Delta_j)
    =
    \exp\!\left[
        -\frac{\alpha\,\Delta_j}
        {\Gamma_{\rm br}+k_{\rm B}T}
    \right].
  \label{eq:Talpha_def}
\end{equation}
Here \(\mathcal W_{0,j}\) is the on-resonance spectral availability of level
\(j\), while \(\Gamma_{\rm br}\) summarizes broadening and inelastic
relaxation. In the simplest implementation \(\mathcal W_{0,j}\) is absorbed
into the prefactor \(I_\ast\), giving
\begin{equation}
    \mathcal W_j^{(\alpha)}
    \simeq
    \exp\!\left[
        -\frac{\alpha\,\Delta_j}
        {\Gamma_{\rm br}+k_{\rm B}T}
    \right].
  \nonumber
\end{equation}

The product \(\alpha\Delta_j\) is therefore the phenomenological
energy-resolved spectral penalty. It parametrizes how strongly the physical
path filters out final many-body states whose spectral weight is displaced
from the tunneling energy. Large \(\alpha\) corresponds to sharp spectral
selectivity. Small \(\alpha\) corresponds to broad spectral availability,
efficient inelastic relaxation, or leakage between nominally distinct
final-state channels. In the limit \(\alpha\to0\), the spectral filter becomes
ineffective and, after absorption of \(\mathcal W_{0,j}\) into \(I_\ast\), the
model approaches
\begin{equation}
    I_t\propto n_E .
  \nonumber
\end{equation}
Thus \(\alpha\) is not a separate transport mechanism. It is the single
phenomenological parameter by which the many-body matrix element and the
energy-resolution factor in Eq.~\eqref{eq:Wj_microscopic} are compressed into a
tractable Landau-level tunneling model.

The tunneling probability contains an overlap factor analogous to a
Franck--Condon factor: the electronic transition is fast compared with the
relaxation of the slow electrostatic or configurational coordinate, so that the
transition amplitude is controlled by the overlap between initial and final
self-consistent configurations.

\subsection{Compressible and locked contributions}
\label{sec:tunnel_LL_sum}

The occupation of the highest Landau level determines whether the
accumulation layer is compressible or locked. If the highest occupied level is
partially filled, removing an electron changes the occupation within a
macroscopically degenerate level. The final state remains on the admissible
physical path and its spectral weight lies at the tunneling energy. Thus
\begin{equation}
    \Delta_{\rm comp}\simeq 0,
    \qquad
    T_\alpha(0)=1.
  \nonumber
\end{equation}
The spectral filter is then open, and the current is governed mainly by the
occupation of the partially filled level selected by the spatial filter.

If the highest occupied level is completely filled, the accumulation layer is
incompressible. Removing one electron would naively produce a compressible
configuration at the same nominal \((n_t,\kappa)\). However, that state is not
necessarily admissible on the physical path. In projector
notation,
\begin{equation}
    \hat P_{\rm phys}(n_t,\kappa)
    |\Psi_{\rm comp}^{N-1}(n_t,\kappa)\rangle
    =
    0 ,
  \nonumber
\end{equation}
so this forbidden branch has zero projected spectral weight and contributes no
current. This is the many-body spectral filtering of the physical path: a
mathematically existing self-consistent configuration does not contribute to
tunneling if it is not an admissible final state of the interacting system.

The remaining tunneling channel is the nearest allowed many-body state. Its
spectral weight can be nonzero, but it is displaced by a protection energy
\begin{gather}
  \Delta_{\rm top}(\kappa)
  = E_{\rm allow}^{N-1}(n_t-\delta n,\kappa)
  - E_{\rm inc}^{N}(n_t,\kappa),
  \nonumber \\
  0 \leq \Delta_{\rm top} \leq \hbar\omega_c.
  \nonumber
\end{gather}
In the charge-partition picture this displacement is controlled by the
screening channel,
\begin{equation}
    \Delta_{\rm top}(\kappa)
    \sim
    \mu_S(n_S).
  \nonumber
\end{equation}
The screening charge is therefore spatially filtered out but spectrally
active: it suppresses tunneling by shifting the allowed many-body spectral
weight away from the tunneling energy.

We use the per-resolved-level occupancy $n_D$
(Sec.~\ref{sec:degeneracy_gap_role}) and denote by \(N_{\rm full}\) the
number of completely filled levels below the partially filled one. The
simulated current is therefore
\begin{equation}
    I_t
    =
    I_\ast
    \sum_j
    n_j\,\mathcal W_j^{(\alpha)}
    =
    I_\ast
    \sum_j
    n_j
    \exp\!\left[
        -\frac{\alpha\,\Delta_j}
        {\Gamma_{\rm br}+k_{\rm B}T}
    \right].
  \nonumber
\end{equation}

For an incompressible configuration, all relevant levels are filled and the
topmost filled level carries the locked-branch spectral penalty:
\begin{equation}
    I_t^{(\rm inc)}
    =
    I_\ast n_D
    T_\alpha(\Delta_{\rm top})
    \sum_{k=0}^{N_{\rm full}-1}
    T_\alpha(k\hbar\omega_c).
  \label{eq:It_inc}
\end{equation}
The first factor describes the displacement of the nearest allowed many-body
final state out of the locked accumulation layer. The geometric series
suppresses tunneling from deeper filled Landau levels.

\begin{figure}
    \centering
    \includegraphics[width=0.95\linewidth]{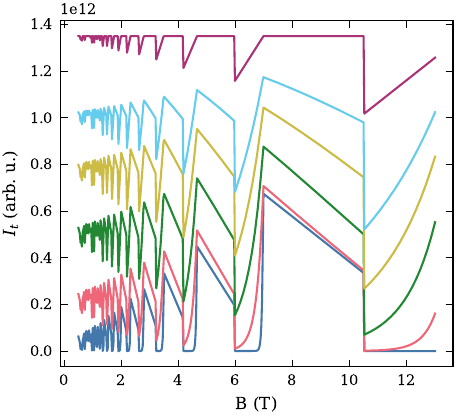}
    \caption{Tunnel current \(I_t\) as a function of magnetic field at fixed
    \(n_t\) for \(\alpha \in [0.3,\,0.05,\,0.02,\,0.01,\,0.005,\,0.0]\),
    \(\gamma=0\), and \(T=0\). The parameter \(\alpha\) controls the sharpness
    of the many-body spectral filter. Large \(\alpha\) suppresses allowed but
    off-resonant final states, leaving mainly the topmost accumulation-layer
    Landau level selected by the spatial filter. In the limit
    \(\alpha\to0\), the spectral penalty disappears and the current approaches
    \(I_t\propto n_E\).}
    \label{fig:tunnel_current_alpha_overlay}
\end{figure}

For a compressible configuration, the topmost level is partially filled and
contributes without the locked-branch penalty:
\begin{equation}
    I_t^{(\rm comp)}
    =
    I_\ast
    \left[
        n_D
        \sum_{k=1}^{N_{\rm full}}
        T_\alpha(k\hbar\omega_c)
        +
        \bigl(n_E-N_{\rm full}n_D\bigr)
    \right].
  \label{eq:It_comp}
\end{equation}
The lower filled levels are suppressed by multiples of \(\hbar\omega_c\). If
\begin{equation}
    r=T_\alpha(\hbar\omega_c),
  \nonumber
\end{equation}
their contribution forms a geometric series. For large \(\alpha\), \(r\ll1\),
and only the topmost available Landau level contributes appreciably. For small
\(\alpha\), \(r\to1\), and the result approaches the total spatially selected
accumulation-layer occupation \(n_E\) (Fig.~\ref{fig:tunnel_current_alpha_overlay}).

For numerical stability near the transition between locked and compressible
segments, we use the smoothed interpolation
\begin{equation}
    I_t
    =
    I_\ast
    \left[
        n_{\rm ex}
        +
        n_D\,T_\alpha(\Delta_{\rm base})
        \sum_{k=0}^{N_{\rm full}-1} r^k
    \right],
    \qquad
    r=T_\alpha(\hbar\omega_c),
  \label{eq:It_unified}
\end{equation}
where
\begin{equation}
    n_{\rm ex}=n_E-N_{\rm full}n_D
  \nonumber
\end{equation}
is the occupation of the partially filled topmost level, and
\begin{equation}
    \Delta_{\rm base}
    =
    f\,\hbar\omega_c
    +
    (1-f)\,\Delta_{\rm top},
    \qquad
    f=
    \min\!\left(
        \frac{n_{\rm ex}}{f_w n_D},
        1
    \right).
  \nonumber
\end{equation}
Here \(f_w\ll1\) is a narrow crossover width. This interpolation is not meant
as a microscopic theory of the crossover; it is a regularized implementation
of the same two-filter model.

\subsection{Comparison with experiment}
\label{sec:comparison_experiment}

Figure~\ref{fig:tunnel_current_alpha_overlay} shows the simulated tunnel
current as a function of magnetic field for several values of $\alpha$. In
the compressible regime, for $\alpha\gtrsim 0.05$, contributions from the
lower filled Landau levels are essentially suppressed and the current
slope directly tracks the linear decrease of the topmost-level occupation
with increasing $B$. At the handover between adjacent filling factors,
$n_E$ readjusts while $\Delta_{\rm top}$ jumps from zero to its
locked-branch value, producing a sudden current drop. In the
incompressible regime, $n_S$ decreases with increasing $B$; because
$T_\alpha$ depends exponentially on $n_S$ through $\Delta_{\rm top}$, the
current recovers rapidly as $n_S\to 0$ and the system re-enters the
compressible state.

\begin{figure}
    \centering
    \includegraphics[width=0.95\linewidth]{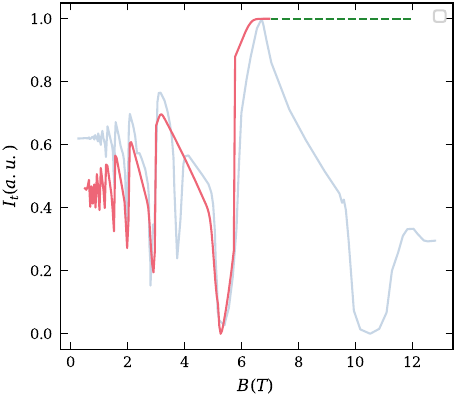}
    \caption{Tunnel current versus magnetic field: experimental data (blue)
    from Ref.~\cite{PhysRevB.38.10120} (sample M5887, $T=1.2\,\mathrm{K}$,
    $V_t=420\,\mathrm{mV}$) compared with the simulation (red). Both curves
    are normalized to their respective peak values. Because the simulation
    does not include spin splitting, the calculated tunnel current saturates
    at a constant value (dotted line) once only the lowest Landau level
    remains occupied at high magnetic fields.}
    \label{fig:exp_It_vs_B_220mV}
\end{figure}

The resulting pattern is a sequence of triangular current structures with
substantial amplitude, directly probing the charge densities in the
compressible and incompressible regimes.
Figure~\ref{fig:exp_It_vs_B_220mV} compares the simulation with the
experimental data of Ref.~\cite{PhysRevB.38.10120}. The approximate
linearity of the current maxima is notable for an intrinsically
exponential process. It arises because the subband energy varies only
weakly with magnetic field, which suppresses the direct exponential field
dependence and leaves the linear occupation dependence as the dominant
modulation. A comparison of Figs.~\ref{fig:tunnel_current_alpha_overlay}
and~\ref{fig:exp_It_vs_B_220mV} confirms the two central
predictions of Eqs.~\eqref{eq:It_inc} and~\eqref{eq:It_comp}:
near-vanishing tunnel current in the incompressible intervals and an
approximately linear increase of the current maxima with magnetic field.

Three remarks on the scope of the present simulation are in order.

First, the equilibrium partition $(n_E,n_S)$ is treated as independent of
temperature and Landau-level broadening. Consequently, a residual
triangular structure survives even in the limit $\alpha\to 0$
[Eq.~\eqref{eq:Talpha_def}], reflecting the sharp staircase in $n_E$.
Physically, the conditions that make $T_\alpha\to 1$ (large $\Gamma_{\rm
br}$, high $T$) would simultaneously round the charge partition. When both
effects are included---a broadened partition together with a reduced
$\alpha$---the residual triangle dissolves smoothly. The restricted
treatment adopted here deliberately decouples these two channels in order
to isolate a logically transparent representation of the tunneling
physics that highlights the relevant mechanisms, namely the observable
consequences of $T_\alpha$. That such a minimal and deliberately
nonoptimized description nevertheless reproduces the dominant qualitative
features of the experimental tunneling data is strong evidence that it
captures the essential underlying mechanism.

Second, Eqs.~\eqref{eq:It_inc} and~\eqref{eq:It_comp} describe the two
extremes of an idealized binary partition. In a physical system the
transition between compressible and incompressible behavior is continuous.
Both limiting cases are encompassed in the unified expression of
Eq.~\eqref{eq:It_unified}, with the interpolated base penalty
$\Delta_{\rm base}=f\,\hbar\omega_c+(1-f)\Delta_{\rm top}$ and crossover
weight $f$ defined there: $\Delta_{\rm base}$ reduces to $\hbar\omega_c$ in
the compressible limit ($f=1$) and to $\Delta_{\rm top}$ in the
incompressible limit ($f=0$), recovering Eqs.~\eqref{eq:It_comp}
and~\eqref{eq:It_inc} respectively. Equation~\eqref{eq:It_unified}
eliminates the binary switch and produces smooth curves when the charge
partition is broadened by finite temperature or disorder. The precise
functional form of the crossover---which encodes the many-body physics of
the compressible--incompressible transition and the microscopic processes
governing charge redistribution into and out of the incompressible
phase---remains an open question, to be addressed by future microscopic
theory and systematic experimental characterization of the transition
region.

Third, only Landau gaps are included in the present calculation. Extending
the same framework to smaller spectral gaps, such as those associated
with spin splitting or fractional quantum Hall states, should carry the
mechanism to lower energy scales and account for the negative triangular
slope observed experimentally for $\nu<1$ at high magnetic fields.

Beyond these scope-related remarks, it is useful to contrast the present
interpretation with the earlier analysis of the same tunneling experiments
in Ref.~\cite{PhysRevB.38.10120}. In that self-consistent description of the
accumulation layer, the boundary condition was chosen so that the Fermi
level remained pinned to the conduction-band edge in the substrate. Within
that picture, increasing Landau-level separation reduces the
two-dimensional carrier density $n_E$, and, together with a weak
modulation of the tunnel coefficient, produces the observed triangular
tunneling characteristic. While that interpretation captures the overall
current profile, it does not account for the large amplitudes at integer
filling or for the pronounced influence of the incompressible regions on
the tunneling signal. In the present framework, by contrast, the boundary
conditions follow from thermodynamic equilibrium of the coupled
subsystems, and the resulting charge-partitioning description provides a
unified account of these additional features.

\section{Discussion}
\label{sec:discussion}
The main result of the paper can now be stated succinctly. Once a spectral
gap prevents the accumulation layer from absorbing further charge
continuously, the heterostructure must store the additional induced charge
and the associated incremental work outside the accumulation layer itself.
The state of the active system is therefore described by the partition
$n_t=n_E+n_S$, a heterostructure free energy is required, and the
physically relevant quantity is the chemical-potential increment
accumulated along the locked branch rather than a pointwise minimization
over $\kappa$.

Within that logic, the roles of the different parts of the theory are
sharply divided. The thermodynamic construction defines how stored work
builds up in the coupled structure. The spectrum decides whether the
system remains on the compressible branch, enters the locked branch, and
when release occurs through $\Delta\mu=\Delta_{\rm gap}$. The microscopic
Poisson--Schr\"odinger analysis then shows where the partitioned charge
lives spatially: a near-interface bound contribution $n_E$ coexists with a
deeper extended screening channel $n_S$. Capacitance and tunneling are
different experimental projections of the same mechanism.

The status of the branch-entry offset deserves a final comment. The
quantity $\mu(1^-,\lam)$ is a real feature of the thermodynamic
construction, and its subtraction is necessary for the definition of the
release potential. A stronger interpretation in terms of nucleation,
latent heat, or protective barriers may well be fruitful, but it should
be presented as an interpretive extension rather than as an established
result of the present work.

\subsection{In-plane scattering}
In-plane scattering obeys the same spectral selection principle as
perpendicular tunneling, but without the spatial barrier factor. For a
scattering potential \(\hat V_{\rm scatt}\), the transition rate is
\begin{equation*}
W_{i\to f}
=
\frac{2\pi}{\hbar}
\left|
\langle \Psi_f|\hat V_{\rm scatt}|\Psi_i\rangle
\right|^2
\delta(E_f-E_i).
\end{equation*}
Thus the scattering cross section is proportional to the final-state spectral
availability,
\begin{equation*}
\sigma_{\rm scatt}
\propto
|V_{\rm scatt}|^2 A_f(E_i).
\end{equation*}
In a compressible Landau level, \(A_f(E_i)\neq0\), and elastic in-plane
scattering is allowed. In an incompressible interval, the Fermi level lies in a
spectral gap, so \(A_f(E_i)=0\) in the ideal limit. Consequently,
\begin{equation*}
\sigma_{\rm scatt}^{\rm inc}=0
\end{equation*}
for elastic scattering at the Fermi level. Inelastic scattering becomes possible
only if phonons, disorder broadening, electron-hole excitations, photons, or
another bath provide spectral weight at the gap energy. With a linewidth
\(\gamma\) and $Z_f$ the spectral weight of the available final state, one
obtains the residual estimate
\begin{equation*}
\sigma_{\rm scatt}^{\rm inc}
\propto
|V_{\rm scatt}|^2 Z_f
\frac{\gamma}{\Delta_{\rm gap}^2+\gamma^2},
\end{equation*}
whereas thermal activation gives
\begin{equation*}
\sigma_{\rm scatt}^{\rm inc}
\propto
|V_{\rm scatt}|^2 Z_f
\exp\!\left[-\frac{\Delta_{\rm gap}}{k_{\rm B} T}\right].
\end{equation*}
Thus the incompressible state suppresses dissipative in-plane scattering not by
removing the scattering potential, but by removing the energetically admissible
final states.

\section{Conclusion}
\label{sec:conclusion}
The central message of this work is that a gapped accumulation layer poses a
charge-redistribution problem for the full heterostructure. Once the
accumulation-layer spectrum has a gap, additional induced charge must be
stored in a complementary screening channel, and the resulting free-energy
increment generates the chemical-potential rise that eventually releases the
system across the relevant gap. This logic motivates the partition
$n_t=n_E+n_S$, the heterostructure free energy, and the branch-resolved
chemical potential in one step.

This framework organizes the analysis of the rest of the paper. The
spectrum selects whether the system remains compressible or evolves along
a locked branch; the locked branch stores the incremental work measured
by $\Delta\mu^*$; release occurs when that increment reaches
$\Delta_{\rm gap}$; and plateau width, capacitance, and tunnel current
emerge as different projections of the same storage-and-release mechanism.
The canonical Poisson--Schr\"odinger reduction makes this framework
calculable across density and geometry, while comparison with experiment
supports the physical picture of a nearby screening reservoir whose active
extent grows with magnetic field.

\subsection*{Connection to quantum-Hall phenomenology.}

Although the present analysis concerns perpendicular transport, it touches a
central ingredient of the integer quantum Hall
effect~\cite{vonKlitzing1980,PrangeGirvin1990}: the existence of finite
intervals in which the electronic system is incompressible. In the present
formulation these intervals correspond to a locked accumulation-layer density,
\begin{equation*}
    n_E=\nu n_D,
\end{equation*}
while the remaining charge is accommodated by the screening channel. This
immediately suggests a connection to the quantized Hall response, since a Hall
measurement governed by \(n_E\) gives
\begin{equation*}
    \rho_{xy}=B/(e n_E)=h/(\nu e^2).
\end{equation*}
The same locked spectral structure suppresses final-state availability for
dissipative scattering, consistent with the vanishing of \(\rho_{xx}\).

We emphasize that this observation is not offered as a complete alternative
theory of the quantum Hall effect. Rather, it points to a possible common
thermodynamic origin for phenomena that are often discussed in different
languages: perpendicular tunneling blockade, capacitance suppression,
integer-filling plateaus, and the in-plane Hall response. Since all of these
effects occur in the same heterostructure and depend on the same
Landau-level incompressibility, it is natural to ask whether the
charge-partition path \(n_t=n_E+n_S\) provides a useful unifying backbone.
Exploring this connection may help bridge perpendicular-probe experiments and
the conventional in-plane quantum-Hall phenomenology. Recent re-examinations
of plateau-width physics~\cite{YiThomas2025} and of the role of disorder in
the quantum-Hall response~\cite{KimKivelson2021} have continued to refine
this picture.

% =====================
% Appendices
% =====================

\clearpage
\appendix

\section{Canonical self-consistent Poisson--Schr\"odinger formulation}
\label{sec:canonical_ps}

\begin{figure}
    \centering
    \includegraphics[width=0.95\linewidth]{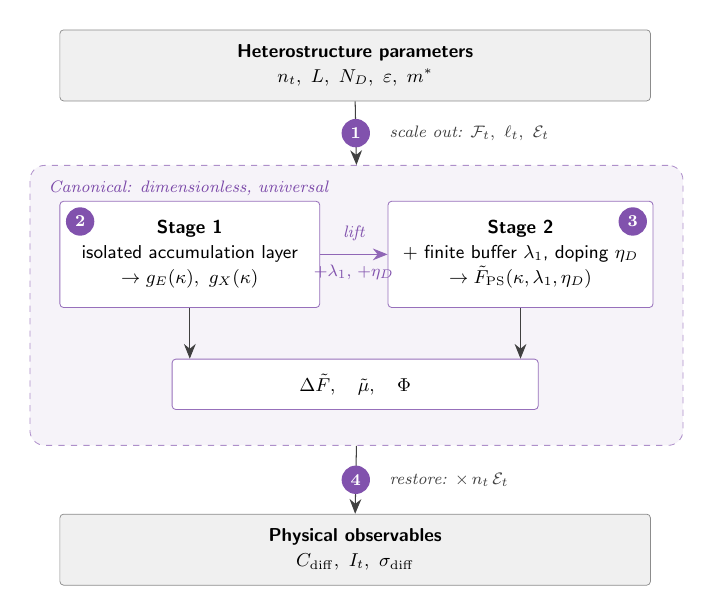}
    \caption{Roadmap of the canonical self-consistent Poisson--Schr\"odinger formulation. Starting from the dimensional heterostructure parameters $(n_t,L,N_D,\varepsilon,m^\ast)$, the problem is first reduced to a dimensionless canonical form by scaling out the natural units $F_t$, $\ell_t$, and $E_t$. In stage 1, the isolated accumulation layer is solved in terms of the one-parameter master functions $g_E(\kappa)$ and $g_X(\kappa)$. In stage 2, this canonical core is lifted to the finite heterostructure by introducing the buffer parameter $\lambda_1$ and doping parameter $\eta_D$, yielding the canonical free-energy and response quantities $\tilde F_{\rm PS}(\kappa,\lambda_1,\eta_D)$, $\Delta\tilde F$, $\tilde\mu$, and $\tilde\Phi$. Dimensional observables such as $C_{\rm diff}$, $I_t$, and $\sigma_{\rm diff}$ are then recovered by restoring the physical scales.}
\label{fig:canonical_bridge}
\end{figure}

This appendix isolates the canonical Poisson--Schr\"odinger (PS) component of the theory and gives it a more explicit internal structure than is possible in the main text. The aim is not merely computational convenience. We show that the self-consistent accumulation problem admits a genuine two-stage canonicalization: first at the level of the isolated accumulation layer, and then again at the level of the finite-buffer multisubband heterostructure. In close analogy with the canonical Airy problem of the triangular well, the result is a reusable set of universal objects that can be tabulated once and then carried into the free-energy, chemical-potential, and release constructions of the main text. A modified Thomas--Fermi treatment of accumulation and depletion layers in semiconductors was developed by Iafrate and Hess~\cite{IafrateHess1989}; the present construction goes beyond Thomas--Fermi by retaining the full self-consistent quantum spectrum but shares the same canonical-scaling logic. Figure~\ref{fig:canonical_bridge} gives a roadmap of this construction relating dimensional inputs to canonical observables.

We take the charge partition and the canonical length and energy scales from the body of the paper, Eqs.~\eqref{eq:nt_partition} and the definitions in Sec.~\ref{sec:device_geometry_complete}, and record them here in the scaled form used throughout this appendix:
\begin{equation}
\begin{aligned}
  \Ft &= \frac{e^2}{\varepsilon}\,\nt,
  \qquad
  \lt = \left(\frac{\hbar^2}{2m^\ast \Ft}\right)^{1/3},
  \\
  \Et &= \Ft\lt,
  \qquad
  \lam = \frac{L}{\lt},
  \label{eq:canonical_scales}
\end{aligned}
\end{equation}
with $L$ the physical width of buffer~1 and $\lam$ its canonical width. For the experimentally relevant densities, $\lt$ is of order $10$~nm, so the interval $\lam\sim 20$ to $100$ already covers the few-hundred-nanometre regime relevant to the data.

\subsection{Overview, motivation, and strategy}
\label{sec:appendixA_overview}

The logic of the appendix is a three-step progression.
\begin{enumerate}
  \item The canonical Airy problem of the triangular well suggests that the accumulation layer should itself admit a scaled, universal description. Stage~1 shows that this remains true after full self-consistent Poisson--Schr\"odinger feedback: the isolated accumulation layer collapses to a one-parameter family on the $(\nt,\kap)$ grid, encoded by master functions rather than by single canonical numbers.
  \item The numerical stage-1 solutions immediately suggest a second step. The bound ground state is sharply localized near the interface, whereas the higher states are orthogonal to it, live at higher energies, and extend deep into the buffer. This motivates splitting the total transferred charge into a bound accumulation-layer component $\kap n_t$ and an extended component $(1-\kap)n_t$ as the natural starting point for a fully self-consistent multisubband treatment.
  \item The central result of stage~2 is that this enlarged problem again admits stable canonical solutions. The finite-buffer multisubband system does not produce a second pair of fitted master curves, but a tabulated family of master surfaces on the $(\kap,\lam)$ grid. These surfaces are the universal input for everything that follows on that grid: the free energy, the chemical potential, and the release map used in the main text.
\end{enumerate}

The appendix is therefore organized as a constructive mathematical-physics narrative. We first state the most general finite-buffer problem in canonical variables, then recover stage~1 as the isolated canonical core, then lift that core to stage~2, and only afterwards analyze the microscopic anatomy, canonical response surfaces, universality, and uniqueness properties of the resulting solution family.

\subsection{Canonical statement of the finite-buffer multisubband problem}
\label{sec:stage2}

The most general canonical formulation of stage~2 is obtained by embedding the accumulation layer into buffer~1 of physical width
\(L\) and canonical width \(\lam=L/\lt\). Buffer~1 is uniformly doped
with donor density \(N_D\), for which the natural dimensionless measure
is
\begin{equation}
  \etaD = \frac{N_D\lt}{\nt}.
  \nonumber
\end{equation}
The physically complete multi-subband PS problem in buffer~1 may then be
written in the compact form
\begin{align}
  -\chi_i'' + u(s)\chi_i &= \varepsilon_i\chi_i,
  \label{eq:stage2_sch_compact}\\
  u''(s) &= \etaD - \sum_i \frac{n_i}{\nt}\,|\chi_i(s)|^2,
  \qquad 0\le s\le \lam.
  \label{eq:stage2_poisson_compact}
\end{align}
This is the form most convenient for both numerics and interpretation.
The occupation numbers separate naturally into the accumulation-layer
charge and the extended sector,
\begin{equation}
  n_0 = \nE = \kap\nt,
  \quad
  \sum_{i\ge 1} n_i = \nS + N_D L
  = \nt\,(1-\kap+\etaD\lam).
  \label{eq:occupation_split}
\end{equation}
The extra \(N_D L\) electrons are the ones required to neutralize the
uniform donor background inside buffer~1.

This point is important for the boundary conditions. Integrating
Eq.~\eqref{eq:stage2_poisson_compact} across buffer~1 gives
\begin{equation}
  u'(\lam)-u'(0) = \etaD\lam - \sum_i \frac{n_i}{\nt}.
  \nonumber
\end{equation}
Using Eq.~\eqref{eq:occupation_split} one finds
\begin{equation}
  \sum_i \frac{n_i}{\nt}=1+\etaD\lam,
  \qquad\Rightarrow\qquad
  u'(\lam)-u'(0)=-1.
  \nonumber
\end{equation}
With the standard interface condition \(u'(0)=1\), the natural
buffer-1 boundary condition is therefore
\begin{equation}
  u'(\lam)=0.
  \nonumber
\end{equation}
In other words, the interface field generated by the transferred charge
is fully screened within buffer~1. This is precisely why buffer~1 is the
correct canonical subsystem for the \(\kap\)-dependent physics. Buffer~2 does
not need to be included at this stage. Its role is only to close the full slab
electrostatically and to contribute a \(\kap\)-independent voltage offset and
capacitance baseline that can be added later.

\subsection{Stage~1: canonical isolated accumulation layer}
\label{sec:stage1}

Stage~1 is the direct self-consistent generalization of the canonical Airy problem. The transferred sheet charge is restricted to the isolated accumulation layer, so the Poisson--Schr\"odinger problem collapses to a one-parameter family indexed by the screening fraction $\kap$. In this sense stage~1 solves the self-consistent accumulation problem once and for all at the level of universal master functions. With the canonical coordinate and potential
\begin{equation}
  s = \frac{x}{\lt},
  \qquad
  u(s) = \frac{U(x)}{\Et},
  \qquad
  \varepsilon_i = \frac{E_i}{\Et},
  \nonumber
\end{equation}
We take $e>0$ and use $U(x)$ for the electron potential energy rather than the
scalar potential, so that $U=-e\phi$. The canonical field $u'(s)$ and the
later gate-voltage formulas are written consistently in this electron-energy
convention. Together with the normalized wavefunctions
\begin{equation}
  \psi_i(x)=\lt^{-1/2}\chi_i(s),
  \qquad
  \int_0^\infty |\chi_i(s)|^2\,\dds = 1,
  \nonumber
\end{equation}
the isolated accumulation layer is governed by
\begin{align}
  -\chi_0'' + u(s)\chi_0 &= \varepsilon_0\chi_0, \nonumber
  \\
  u''(s) &= -\kap |\chi_0(s)|^2.
  \label{eq:stage1_poisson}
\end{align}
The natural boundary conditions are
\begin{equation}
  u(0)=0,
  \qquad
  u'(0)=1,
  \qquad
  u'(\infty)=1-\kap.
  \label{eq:stage1_bc}
\end{equation}
Equation~\eqref{eq:stage1_bc} expresses the physical meaning of
\(\kap\): the bound accumulation layer screens only the fraction
\(\kap\) of the interface field, so that the unscreened residual field at
large distance is \(1-\kap\).

\begin{figure}
    \centering
    \includegraphics[width=0.95\linewidth]{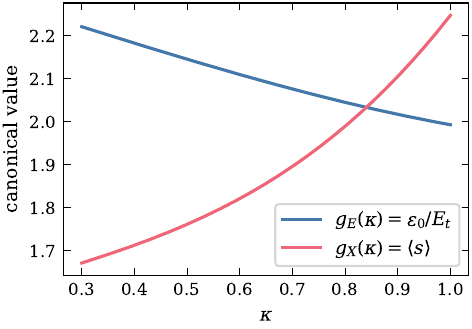}
    \caption{Stage-1 master functions of the canonical isolated-accumulation-layer problem. Blue: dimensionless ground-state energy $g_E(\kappa)=\varepsilon_0/E_t$. Red: dimensionless charge centroid $g_X(\kappa)=\langle s\rangle$. The smooth curves are polynomial fits (poly3) to the numerical solutions. Both master functions vary monotonically with the partition parameter $\kappa$, interpolating between the fully screened limit $\kappa=1$ and the vanishing-bound-charge limit $\kappa\to 0$.}
\label{fig:master_functions}
\end{figure}

In physical units the problem depends on the pair \((\nt,\kap)\),
because \(\lt\) and \(\Et\) depend on \(\nt\). After the scaling in
Eq.~\eqref{eq:canonical_scales}, however, the \emph{shape} of the
self-consistent solution depends only on \(\kap\). Thus the exact PS
problem defines a universal family of canonical solutions
\begin{equation}
  \chi_0(s;\kap),
  \qquad
  u(s;\kap),
  \qquad
  \varepsilon_0(\kap).
  \nonumber
\end{equation}
The corresponding physical observables are obtained by restoring the
canonical scales,
\begin{equation}
\begin{aligned}
  E_0(\nE,\nS)&=\Et(\nt)\,\gE(\kap),
  \\
  \langle x\rangle(\nE,\nS)&=\lt(\nt)\,\gX(\kap).
  \label{eq:master_statements}
\end{aligned}
\end{equation}
Here the master functions are defined by the gauge-invariant canonical
combinations
\begin{equation}
  \gE(\kap)=\varepsilon_\kap-u_\kap(0),
  \qquad
  \gX(\kap)=\int_0^\infty s\,\chi_\kap(s)^2\,\dds.
  \label{eq:gEgx_defs}
\end{equation}
Although the interface gauge is fixed in the present implementation by
$u(0)=0$, the combination $\varepsilon-u(0)$ is retained in order to make the
master-function definition manifestly gauge-invariant and directly transferable
to alternative gauges or finite-barrier variants.
After solution of the
self-consistent PS equations, Eqs.~\eqref{eq:master_statements} and
\eqref{eq:gEgx_defs} become concrete numerical objects: one
may tabulate them, interpolate them, or fit them by simple analytic
forms. In that sense the Airy problem yields canonical numbers, whereas
the self-consistent accumulation-layer problem yields canonical
functions. These master functions are shown in Fig.~\ref{fig:master_functions}.

The stage-1 scaling may also be written explicitly as
\begin{equation}
  E_0(\nt,\kap)=C_E\,\nt^{2/3}\,\gE(\kap),
  \quad
  C_E=\left(\frac{\hbar^2}{2m^\ast}\right)^{1/3}
       \left(\frac{e^2}{\varepsilon}\right)^{2/3},
  \nonumber
\end{equation}
with the centroid scaling
\begin{equation}
  \langle x\rangle(\nt,\kap)=C_x\,\nt^{-1/3}\,\gX(\kap),
  \qquad
  C_x=\left(\frac{\hbar^2\varepsilon}{2m^\ast e^2}\right)^{1/3}.
  \nonumber
\end{equation}
The ratio
\begin{equation}
  R(\kap)\equiv\frac{E_0}{\Ft\langle x\rangle} = \frac{\gE(\kap)}{\gX(\kap)}
  \nonumber
\end{equation}
provides a compact consistency check on the numerics.

\subsubsection{Canonical limiting cases and checks}
The familiar triangular well is recovered at \(\kap=0\). Then
Eq.~\eqref{eq:stage1_poisson} reduces to \(u''=0\), so the potential
is linear and the ground state is the Airy problem. One obtains the
exact reference values
\begin{equation}
\begin{aligned}
  \gE(0)&=|a_1|\simeq 2.3381,
  \qquad
  \gX(0)=\frac{2}{3}|a_1|\simeq 1.5587,
  \\
  R(0)&=\frac{3}{2}.
  \nonumber
\end{aligned}
\end{equation}
At the opposite end, \(\kap=1\) corresponds to the fully screened
self-consistent accumulation layer. This slice is every bit as canonical
as the Airy slice, but it is usually not recognized as such because no
closed form exists. Numerically one finds for the representative fit
used here
\begin{equation}
  \gE(1)\simeq 1.993,
  \qquad
  \gX(1)\simeq 2.247,
  \nonumber
\end{equation}
with the full interval \(0\le \kap\le 1\) interpolating continuously
between the unscreened Airy problem and the fully screened
self-consistent PS problem. A further numerical/canonical observation of the
fully screened slice is that it supports only one bound state in the
self-consistent PS problem; higher eigenvalues belong to the finite-width shelf
and become box-like rather than genuinely bound.

\subsubsection{Numerical solution and master-function representation}
In the present implementation the exact self-consistent solver was run on
a grid of \(\kap\)-values and the resulting canonical curves were fitted
by third-order polynomials,
\begin{equation}
  g(\kap)\approx a+b\kap+c\kap^2+d\kap^3.
  \nonumber
\end{equation}
For \(\gE(\kap)\) we use
\begin{equation}
  \gE(\kap) \approx
  2.3361 - 0.37532\,\kap - 0.056961\,\kap^2 + 0.089119\,\kap^3,
  \label{eq:gE_fit}
\end{equation}
and for \(\gX(\kap)\)
\begin{equation}
  \gX(\kap) \approx
  1.5544 + 0.44947\,\kap - 0.38462\,\kap^2 + 0.62792\,\kap^3.
  \nonumber
\end{equation}
The constant term in Eq.~\eqref{eq:gE_fit} reproduces the Airy value
\(|a_1|\) within numerical tolerance, as it must, and the ratio
\(\gE(0)/\gX(0)\) reproduces the Airy virial value \(3/2\).

\subsection{Stage~2: finite-buffer lift with extended states}
\label{sec:stage2_lift}

With the full finite-buffer boundary-value problem now stated, the second canonicalization consists in lifting the stage-1 accumulation-layer solution into the multisubband problem on the finite interval $0\le s\le \lam$. The logic mirrors stage~1: one fixes the bound component by $n_0=\kap n_t$, distributes the remaining charge over the extended manifold, and searches for a self-consistent potential that closes the Poisson and Schr\"odinger equations on the same canonical grid. The novelty is that the output is no longer a one-parameter family of curves, but a two-parameter family of tabulated surfaces.

\subsubsection{Extended-state occupation and jellium compensation}
\label{sec:neutral_base_map}

Although Eqs.~\eqref{eq:stage2_sch_compact} and
\eqref{eq:stage2_poisson_compact} are the complete buffer-1
formulation, the canonical \(\kap\)-dependent physics becomes clearer
after separating the locally neutral jellium background from the
interfacial residual charge. The donor background and its neutralizing
extended electrons form a nearly field-free subsystem. What remains is
the residual interfacial problem,
\begin{equation}
  u''(s) = -\kap |\chi_0(s)|^2 - \rho_{\rm screen}(s),
  \label{eq:neutral_poisson}
\end{equation}
with
\begin{equation}
\begin{aligned}
  \int_0^\lam \rho_{\rm screen}(s)\,\dds &= 1-\kap,
  \qquad
  u(0)=0,
  \\
  u'(0)&=1,
  \qquad
  u'(\lam)=0.
  \nonumber
\end{aligned}
\end{equation}
Equation~\eqref{eq:neutral_poisson} is the effective
\(\etaD=0\) base map used for the canonical free-energy and release
calculations: all charges that remain in the equation sum to unity, so
the field drops from \(u'(0)=1\) to \(u'(\lam)=0\) entirely within
buffer~1.

The important logical point is that Eqs.~\eqref{eq:stage2_poisson_compact}
and \eqref{eq:neutral_poisson} are not rival descriptions. The former
is the physically complete compact Poisson equation for buffer~1; the
latter is the reduced canonical equation for the \(\kap\)-dependent
residual after the neutral jellium has been factored out.

\subsubsection{Numerical solution and self-consistent map production}
\label{sec:stage2_solver}

The finite-buffer lift becomes useful only if the enlarged multisubband problem can again be solved reproducibly on a canonical grid. The buffer-1 problem is therefore solved self-consistently on the finite interval
\(0\le s\le \lam\) by finite-difference discretization of
Eqs.~\eqref{eq:stage2_sch_compact} and
\eqref{eq:stage2_poisson_compact}. For a given pair
\((\kap,\lam)\), the bound occupation is fixed by construction,
\begin{equation}
  n_0=\kap\,\nt,
  \qquad
  n_{\rm ext}=(1-\kap)\nt,
  \nonumber
\end{equation}
while the extended component is distributed over the subbands
\(i\ge 1\) by Fermi filling. Numerically this is done by solving for the
auxiliary chemical potential that enforces the required extended charge,
\begin{equation}
  \sum_{i\ge 1} n_i(\mu_{\rm ext}) = n_{\rm ext}+N_D L,
  \label{eq:root_fill}
\end{equation}
using a one-dimensional root finder (Brent's method in the present
implementation).

A typical self-consistency cycle proceeds as follows.
\begin{enumerate}
  \item Start from an initial guess \(u^{(0)}(s)\), typically the linear
  triangular profile or the converged solution at a nearby map point.
  \item Diagonalize the discretized Schr\"odinger operator to obtain
  \(\varepsilon_i^{(m)}\) and \(\chi_i^{(m)}\) at iteration \(m\).
  \item Determine the extended occupations \(n_i^{(m)}\) from
  Eq.~\eqref{eq:root_fill}.
  \item Build the total electronic density and solve Poisson's equation
  for the updated potential.
  \item Apply clamped linear mixing,
  \begin{equation}
    u^{(m+1)} \leftarrow
    (1-\alpha)u^{(m)}+\alpha\,u_{\rm new}^{(m)},
    \qquad 0<\alpha\ll 1,
  \nonumber
\end{equation}
  to suppress oscillatory convergence in the high-\(\kap\) regime.
\end{enumerate}
For the dense production runs used here, convergence typically requires
of order \(10^2\) iterations, with representative difficult points near
\(\kap\to 1\) taking roughly \(\sim 200\) iterations.

\begin{figure}
    \centering
    \includegraphics[width=0.95\linewidth]{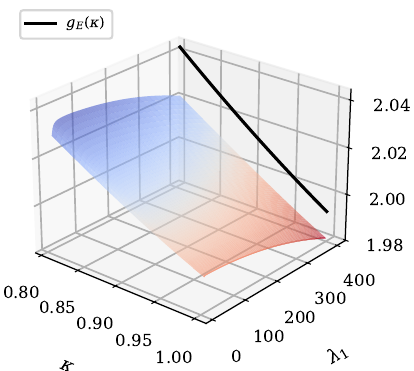}
    \caption{Stage-2 canonical ground-state energy $\varepsilon_0(\kappa,\lambda_1)$ obtained from the self-consistent finite-buffer map on a $26\times 98$ grid in $(\kappa,\lambda_1)$. The color surface (coolwarm) shows the dimensionless energy landscape, while the black line at $\lambda_{\max}$ indicates the stage-1 asymptote $g_E(\kappa)$. For small $\lambda_1$, the energy is increased by finite-size quantization; with increasing $\lambda_1$, the surface approaches the stage-1 master function, demonstrating the recovery of the isolated-accumulation-layer limit.}
\label{fig:gE_surface}
\end{figure}

The resulting canonical map is produced on a dense rectangular grid in
\((\kap,\lam)\). A representative production map uses
\begin{equation}
  \kap\in[0.30,1.00],
  \qquad
  \lam\in[10,400],
  \qquad
  N_{\kap}\times N_{\lam}=150\times 75,
  \nonumber
\end{equation}
corresponding to \(11\,250\) self-consistent solves. Two regimes emerge
clearly from this map. In Regime~I, \(\kap\in[0.80,1]\), all relevant
observables vary smoothly with \((1-\kap)\) and admit the fitted forms
quoted above. In Regime~II, \(\kap<0.80\), discrete subband filling
introduces visible kinks and non-smooth structure. The application of
interest lies entirely in Regime~I, which is why the stage-2 master
functions acquire such a compact representation there.
Figure~\ref{fig:gE_surface} shows the resulting stage-2 surface
$g_E(\kappa,\lambda_1)$ and its approach to the stage-1 asymptote at
large $\lambda_1$.

A technical issue occurs exactly at \(\kap=1\). The purely bound,
single-component state has no extended occupation, so some derivatives
of the finite-region free energy exhibit a one-sided discontinuity at the
onset of the two-component branch. In practice this is regularized for
interpolation by replacing the map value at \(\kap=1\) with a linear
extrapolation from nearby points on the locked branch, while storing the
true compressible value separately. This is the numerically stable way to
construct the spline derivatives needed later for \(\Delta\mu^\ast\).

\subsection{Microscopic anatomy of the stage-2 solution}
\label{sec:microscopic_anatomy}

The canonical stage-2 map is more than a repository of fitted functions.
It also exposes the internal structure of the self-consistent buffer
solution and thereby gives direct microscopic meaning to the thermodynamic
partition of Eq.~\eqref{eq:nt_partition}. The accumulated
component \(\nE\) is carried by the interfacial ground state and remains
concentrated in the near-interface core. The extended component \(\nS\) is
built from the higher subbands and spreads across the buffer. Three
observations make this separation visible: the orthogonality minimum of the
excited states, the incoherent Fermi-weighted sum of those states, and the
corresponding decomposition of the screening field.

\subsubsection*{Orthogonality minimum}

The first excited and higher subbands are constrained by orthogonality to
the accumulated ground state,
\begin{equation}
  \int_0^{\lam} \chi_0(s)\chi_i(s)\,\dds = 0,
  \qquad i\ge 1.
  \nonumber
\end{equation}
Because \(\chi_0\) is positive and strongly concentrated near the
interface, every higher state must suppress its weight in that same core
region. As a result, the excited-state densities develop a pronounced
minimum, and in practice often a node, close to the interface. Denoting
this position by \(s_{\rm min}^{(i)}\), one finds over the shelf-like part
of the self-consistent potential the approximate collapse law
\begin{equation}
  s_{\rm min}^{(i)} \simeq \frac{\pi}{\sqrt{\varepsilon_i}},
  \nonumber
\end{equation}
which shows that the onset of the extended sector is not accidental but a
direct consequence of orthogonality within the canonical PS spectrum. In
this sense the higher states are structurally expelled from the
interfacial accumulation core.

\subsubsection*{Incoherent sum and Fermi-weighted density}

The higher-subband sector is therefore not represented by a single state. Its
full Fermi-weighted electronic density in buffer~1 is
\begin{equation}
  \rho_{\rm ext}^{\rm full}(s;\kap,\lam)=\sum_{i\ge 1}\frac{n_i}{\nt}\,|\chi_i(s)|^2,
  \nonumber
\end{equation}
which contains both the electrons required to neutralize the donor background
and the residual compensating charge associated with the partition variable
$\nS$. The accumulated component is
\begin{equation}
  \rho_E(s;\kap,\lam)=\kap\,|\chi_0(s)|^2.
  \nonumber
\end{equation}
To isolate the genuinely compensating density that enters the reduced canonical
problem, we subtract the neutral jellium background and define
\begin{equation}
  \rho_S(s;\kap,\lam)=\rho_{\rm ext}^{\rm full}(s;\kap,\lam)-\etaD,
  \nonumber
\end{equation}
so that
\begin{equation}
  \int_0^\lam \rho_S(s;\kap,\lam)\,\dds = 1-\kap.
  \nonumber
\end{equation}
The total residual electronic density in the canonical buffer problem is then
\begin{equation}
  \rho_{\rm tot}(s;\kap,\lam)=\rho_E(s;\kap,\lam)+\rho_S(s;\kap,\lam).
  \nonumber
\end{equation}
At \(\kap=1\), one has \(\rho_S=0\) in this reduced residual sense and the
solution is purely the accumulated interfacial state. For \(\kap<1\), the
excited subbands acquire finite occupation through Fermi filling, and their
higher-subband density splits naturally into a neutral jellium part and a broad
residual screening profile across buffer~1. The thermodynamic decomposition
into \(\nE\) and \(\nS\) is therefore not an abstract bookkeeping device; it is
the energetic representation of two self-consistently distinct spatial sectors
of the canonical buffer solution.

\subsubsection*{Screening decomposition}

The same separation appears directly in Poisson's equation. In the full
buffer-1 formulation one has
\begin{equation}
  u''(s)=\etaD-\rho_E(s;\kap,\lam)-\rho_{\rm ext}^{\rm full}(s;\kap,\lam),
  \nonumber
\end{equation}
whereas in the reduced residual-field formulation of
Eq.~\eqref{eq:neutral_poisson} this becomes
\begin{equation}
  u''(s)=-\rho_E(s;\kap,\lam)-\rho_S(s;\kap,\lam).
  \nonumber
\end{equation}
It is useful to compare the full self-consistent field to the field that
would be generated by the accumulated sector alone,
\begin{equation}
  u_E''(s)=-\rho_E(s;\kap,\lam),
  \qquad
  u_E'(0)=1,
  \nonumber
\end{equation}
with the same interface normalization. Near the interface the bound sector
controls the field, and the behavior is close to the purely accumulated
solution. Deeper into the buffer, however, the extended sector provides
both the donor-neutralizing higher-subband background and the decisive residual
nonlocal screening channel, bending the field away from the bound-only result.
The canonical buffer solution therefore has a genuine two-regime structure: a
near-interface accumulated core and an extended screening tail. This picture is
the microscopic origin of the later voltage redistribution, capacitance
response, and tunneling blockade.

\subsection{Canonical response surfaces and interface--bulk decomposition}
\label{sec:stage2_master_objects}

The stage-2 self-consistent map on the $(\kap,\lam)$ grid is valuable not only because it solves the finite-buffer problem numerically, but because it organizes the result into a small set of canonical response surfaces. These outputs separate into near-interface quantities, which are controlled locally by the bound state and saturate rapidly with $\lam$, and bulk quantities, which retain the genuine finite-width dependence of the extended sector.

\subsubsection{Definitions on the $(\kap,\lam)$ grid}

The heterostructure-corrected ground-state master functions are
defined, by analogy with their stage-1 counterparts, as
gauge-invariant combinations of the converged stage-2 solution:
\begin{equation}
\begin{aligned}
  \gEH(\kap,\lam)
  &= \varepsilon_0(\kap,\lam)-u(0;\kap,\lam),
  \\
  \gXH(\kap,\lam)
  &= \int_0^\lam s\,|\chi_0(s;\kap,\lam)|^2\,\dd s,
  \nonumber
\end{aligned}
\end{equation}
As in stage~1, the subtraction of $u(0)$ is retained to make the definition
manifestly gauge-invariant, even though the numerical implementation adopts the
interface gauge $u(0)=0$. The physical observables are then recovered by the
canonical scales:
\begin{equation}
\begin{aligned}
  E_0^{(H)}(\nt,\kap,\lam) &= \Et(\nt)\,\gEH(\kap,\lam),
  \\
  \langle x\rangle^{(H)}(\nt,\kap,\lam)
  &= \lt(\nt)\,\gXH(\kap,\lam).
  \nonumber
\end{aligned}
\end{equation}
They decompose into the stage-1 master functions plus an additive
Hartree correction:
\begin{equation}
  \gEH = \gE(\kap) + \delta g_E^{(H)}(\kap,\lam),
  \quad
  \gXH = \gX(\kap) + \delta g_X^{(H)}(\kap,\lam).
  \nonumber
\end{equation}
The stage-2 map also yields the buffer-1 voltage drop and field
energy,
\begin{equation}
\begin{aligned}
  \Delta\tilde{V}_1(\kap,\lam)
  &= \int_0^\lam u'(s;\kap,\lam)\,\dd s,
  \\
  \mathcal{I}_1(\kap,\lam)
  &= \tfrac{1}{2}\int_0^\lam [u'(s;\kap,\lam)]^2\,\dd s,
  \nonumber
\end{aligned}
\end{equation}
both of which are functions of $(\kap,\lam)$. All five
quantities---$\gEH$, $\gXH$, $\Delta\tilde{V}_1$,
$\mathcal{I}_1$, and the screening ratio $S$---are computed on
the same $(\kap,\lam)$ grid.

\subsubsection{Physical separation: why the ground-state
  corrections lose their $\lam$-dependence}

Although all stage-2 objects are formally defined on
$(\kap,\lam)$, the Hartree corrections $\delta g_E^{(H)}$ and
$\delta g_X^{(H)}$ turn out to be $\lam$-independent in the
regime of interest. This is not an approximation; it reflects a
physical separation between two classes of quantities.

The ground-state energy $\varepsilon_0$ and centroid
$\langle s\rangle_0$ are properties of $\psi_0$, which is
exponentially localized within $\sim 1/b = \gX/3 \approx
0.75\,\lt$ of the interface. For any well width satisfying
$\lam \gg 1/b$---a condition met for $\lam \gtrsim 10$, always
in the regime of interest---the far boundary at $s = \lam$ is
invisible to $\psi_0$. The Hartree corrections to $\varepsilon_0$
and $\langle s\rangle_0$ arise from two local mechanisms: the
jellium mean-field potential (parametrized by $\etaD$), and the
feedback from extended-state screening on the near-interface
potential (parametrized by $\kap$). Both are near-interface
effects that saturate once $\lam$ exceeds the ground-state
localization length.

By contrast, the voltage drop $\Delta\tilde{V}_1$, the field
energy $\mathcal{I}_1$, and the extended kinetic energy
$E_{\rm ext}$ involve the spatial structure of the charge
distribution across the entire well. The extended eigenenergies
scale as $\varepsilon_i \propto i^2/(\lam-s_0)^2$; the field
energy scales as $\mathcal{I}_1 \propto \lam^{0.51}$. These
quantities depend genuinely on $\lam$.

The three canonical variables thus separate by physical origin. The
partition coordinate $\kap$ governs the charge partition between bound
and extended sectors and enters every quantity. The doping parameter
$\etaD = N_D\lt/\nt$ describes the jellium perturbation of the
near-interface potential and enters $\gEH$, $\gXH$, and $\Delta F$
analytically. The buffer width $\lam = L/\lt$ sets the extended-state
spectrum and spatial extent and enters $\Delta F$, $\Delta\mu$,
$\Delta\tilde{V}_1$, and $\mathcal{I}_1$.

\subsubsection{Fitted Hartree-corrected master functions:
  $(\kap,\etaD)$}

The $\lam$-independence of the ground-state corrections allows
$\gEH$ and $\gXH$ to be parametrized as functions of
$(\kap,\etaD)$ alone. In the regime $\kap \in [0.8, 1]$,
$\etaD \in [0, 0.74]$, the numerical data are represented by
\begin{align}
  \gEH(\kap,\etaD)
  &= \gE(\kap)
    + \underbrace{1.157\,\etaD\vphantom{\big|}}_{\text{jellium uplift}}
  \nonumber \\
  &\quad + \underbrace{(-0.033 - 0.095\,\etaD)\vphantom{\big|}}_{\text{screening feedback}}\,(1{-}\kap),
  \nonumber \\
  \gXH(\kap,\etaD)
  &= \gX(\kap)
    + \underbrace{(-0.957\,\etaD)\vphantom{\big|}}_{\text{jellium compression}}
  \nonumber \\
  &\quad + \underbrace{(0.077 + 0.236\,\etaD)\vphantom{\big|}}_{\text{screening feedback}}\,(1{-}\kap).
  \nonumber
\end{align}
Each formula has three parts: the bare stage-1 master function
$\gE(\kap)$ or $\gX(\kap)$; a $\kap$-independent doping shift
(present even at $\kap = 1$, where no extended states exist);
and a $(1{-}\kap)$-dependent correction from extended-state
screening feedback, which vanishes at $\kap = 1$ as it must.

The doping shifts have transparent physical content. The
$+1.157\,\etaD$ in $\gEH$ is the jellium mean-field uplift of
$\varepsilon_0$: the uniform positive background raises the
potential energy at the interface. The $-0.957\,\etaD$ in
$\gXH$ is the corresponding compression of $\psi_0$ toward
the interface: the steeper potential confines the charge more
tightly. The $(1{-}\kap)$ slopes describe how the onset of
extended-state occupation modifies the near-interface
self-consistent potential: filling extended states slightly
lowers $\varepsilon_0$ (additional screening of the interface
field) and pushes the centroid outward.

\subsubsection{Fitted bulk quantities: $(\kap,\lam)$ and
  $(\kap,\etaD)$}

The remaining stage-2 objects retain their $\lam$-dependence.
The screening ratio and exchange voltage, fitted across the
$(\kap,\etaD)$ plane, take the power-law forms
\begin{align}
  1 - S(\kap,\etaD)
  &\approx 0.0182\,\etaD^{-0.92}\,(1{-}\kap)^{0.578}, \nonumber
  \\
  \Delta V_x / \Et
  &\approx (43.1 + 7.3\,\etaD)\,(1{-}\kap)^{0.922}. \nonumber
  \end{align}
These are $\lam$-independent to the accuracy of the fit
because the screening ratio is a near-interface quantity
(the field profile normalized to its unscreened value,
evaluated in the region where $\psi_0$ dominates).

The voltage drop across buffer~1, by contrast, depends on
$\lam$ because it integrates the field across the entire
well. At the reference width $\lam = 54$:
\begin{equation}
  \Delta\tilde{V}_1(\kap;\lam{=}54)
  \approx 2.247 + 5.176\,(1{-}\kap)^{0.605}.
  \nonumber
\end{equation}
The baseline $2.247 = \gX(1)$ is the shelf height
($\lam$-independent for $\lam \gg 1/b$); the correction
amplitude $5.176$ carries implicit $\lam$-dependence through
the extent of the nonlocal screening charge. For the free
energy, this $\lam$-dependence is captured explicitly by the
factor $(1 - 0.756/\lam^{0.163})$ in
Eq.~\eqref{eq:lambda_factor}.

\subsubsection{Computational consequence}

The near-interface / bulk separation has a direct
computational benefit: the self-consistent map need only be
computed on a two-dimensional $(\kap,\lam)$ grid at
$\etaD = 0$ (the neutral effective formulation of
Sec.~\ref{sec:neutral_base_map}). The doping correction
enters through $\gEH$ and $\gXH$ as a separable analytical
perturbation $\propto \etaD$, not through a second numerical
sweep. This reduces the problem from a three-dimensional
parameter space to a two-dimensional map supplemented by
closed-form corrections---a reduction that follows from the
exponential
localization of the ground state.

\subsubsection{Stage-2 free-energy decomposition}
\label{sec:stage2_free_energy}

One reason the stage-2 map is so valuable is that it wraps up the entire
canonical PS chapter in the precise form needed by the thermodynamic
theory. The dimensionless Kohn--Sham free energy of buffer~1 is written
as
\begin{equation}
  \tilde F(\kap,\lam)=\kap\,\varepsilon_0 + E_{\rm ext} - E_H,
  \nonumber
\end{equation}
with
\begin{equation}
  E_{\rm ext}=\sum_{i\ge 1}\frac{n_i}{\nt}\,\varepsilon_i,
  \qquad
  E_H=\frac12\int_0^\lam [u'(s)]^2\,\dds.
  \nonumber
\end{equation}
The three terms have a direct physical meaning: the bound-state
contribution, the energetic cost of populating the extended manifold,
and the Hartree saving associated with electrostatic self-consistency.

In Regime~I, the three contributions follow remarkably simple power laws.
For the canonical base map one finds
\begin{align}
  \Delta\bigl(\kap\varepsilon_0\bigr)
  &\approx -2.297(1-\kap)^{1.180}, \nonumber
  \\
  E_{\rm ext}
  &\approx 3.847(1-\kap)^{1.086}, \nonumber
  \\
  \Delta E_H
  &\approx 0.509(1-\kap)^{0.809}, \nonumber
  \end{align}
so that the canonical free-energy balance is already visible at the PS
level. The finite-width dependence enters to excellent approximation as a
multiplicative factor,
\begin{equation}
  \tilde F(\kap,\lam)
  \propto (1-\kap)^p\left(1-\frac{0.756}{\lam^{0.163}}\right),
  \label{eq:lambda_factor}
\end{equation}
and the explicit donor correction is
\begin{equation}
  \delta\tilde F_D
  \approx -\etaD(1-\kap)\bigl[1.157+0.095\kap\bigr].
  \nonumber
\end{equation}
This is the precise sense in which the stage-2 map prepares the later
thermodynamic treatment: the same canonical objects that describe the
self-consistent confinement problem also provide the compact energetic
building blocks used in the free-energy, chemical-potential, and release
analysis.

\subsection{Canonical character and universality}
\label{sec:why_canonical}

The formulation is canonical in the following precise sense.
\begin{enumerate}
  \item The isolated accumulation layer defines a universal family of
  self-consistent solutions on the \((\nt,\kap)\) grid, all dependence on
  \(\nt\) being carried by the similarity scales \(\lt\) and \(\Et\).
  \item Buffer~1 then promotes this local family to a finite-region map
  on the \((\kap,\lam)\) grid, with \(\lam\) the canonical width of the
  region in which the residual field is screened.
  \item Buffer~2 and the barrier are excluded at this stage, and the
  quasimetallic electrodes enter only as external boundaries, because these
  smooth closure elements do not modify the canonical \(\kap\)-dependent
  physics; they only shift the absolute measured voltage and the
  capacitance baseline.
\end{enumerate}
The canonical PS problem is therefore a transferable mathematical
structure rather than a device-specific numerical exercise. The Airy slice
\((\kap=0)\), the fully screened self-consistent slice \((\kap=1)\), and
the entire interval \(0<\kap<1\) belong to the same canonical family.
The stage-1 master functions \(\gE\) and \(\gX\) and the stage-2
heterostructure-corrected functions \(\gEH\) and \(\gXH\) are the
concrete numerical objects that carry this structure into the
thermodynamic theory.

\subsection{Uniqueness, stability, and limits of the stage-2 lift}
\label{sec:kappa1_one_bound_state}

A foundational structural property of the fully screened canonical
self-consistent Poisson--Schr\"odinger problem is that the
\emph{stage-1 accumulation-layer potential at $\kap=1$ supports exactly
one bound eigenstate}. This is the configuration in which the entire
transferred sheet density resides in the accumulation-layer sector, the
interface field is fully screened, and the self-consistent potential
rises from the interface ramp to a finite shelf value
\(u_\infty=g_X(1)\). The ground state lies below this shelf, whereas all
higher eigenstates lie above it and therefore belong to the extended
sector.

This is the canonical spectral property that makes the two-population
decomposition mathematically sharp at the fully screened endpoint: one
bound state carries the accumulation-layer charge, and the remainder of
the spectrum forms the extended manifold. In particular, it justifies the use of
\(\kap\) as the fractional occupation of the unique bound sector and
shows that the decomposition into ``bound'' and ``extended'' channels is
already present in the self-consistent confinement problem itself.

The argument proceeds by rewriting the fully screened problem as a bound-state
counting problem on the half-line. At \(\kap=1\), the canonical
self-consistent equations are
\begin{subequations}
\begin{align}
  -\chi''(s)+u(s)\chi(s) &= \varepsilon\,\chi(s), \nonumber
  \\
  u''(s) &= -|\chi(s)|^2, \nonumber
  \end{align}
\end{subequations}
with
\begin{equation}
\begin{aligned}
  \chi(0)=0,
  \qquad
  \chi(\infty)&=0,
  \qquad
  u'(0)=1,
  \qquad
  u'(\infty)=0,
  \\
  \int_0^\infty |\chi(s)|^2\,\dds&=1.
  \nonumber
\end{aligned}
\end{equation}
The self-consistent solution rises linearly near the interface and then
saturates to a finite shelf. Defining the effective well depth below the
shelf by
\begin{equation}
  W(s) \coloneqq u_\infty-u(s) \ge 0,
  \nonumber
\end{equation}
all bound states of the original problem correspond to bound states of
\(W(s)\) on the half-line with Dirichlet boundary conditions. The
question ``How many bound accumulation-layer states exist at
\(\kap=1\)?'' therefore reduces to the spectral counting problem for the
effective well \(W(s)\).

Several independent tests have been carried out for this problem. The
Bargmann--Schwinger integral
\begin{equation}
  B=\int_0^\infty s\,W(s)\,\dds
  \nonumber
\end{equation}
and the corresponding Calogero functional are both too loose to prove
uniqueness for the present well: they overcount because the
self-consistent well is deep enough that simple integral criteria do not
exclude a second state. This failure is not a weakness of the canonical
solution, but rather a reflection of the fact that these universal
bounds are insensitive to the detailed shape of a well that is deep but
very narrow.

The decisive shape-sensitive criterion is the semiclassical phase
integral
\begin{equation}
  \Phi=\int_0^\infty \sqrt{W(s)}\,\dds.
  \nonumber
\end{equation}
For a second bound state to exist on the half-line, one requires
\(\Phi\ge 2\pi\). For the converged canonical self-consistent
\(\kap=1\) potential, however, the phase integral remains well below
this threshold. Physically, the well is deep enough to sustain the
single ground-state oscillation, but too narrow to accommodate the two
oscillations required of a second bound state. This is exactly the
spectral fingerprint of self-consistent screening: the same ground-state
charge that creates the confining well also sharpens it and limits its
width.

This conclusion is then confirmed directly by numerical diagonalization
of the effective well on successively refined grids. Across the
converged range, only the ground-state eigenvalue lies below the shelf;
the next eigenvalue already lies slightly above it. In other words, the
second state is not strongly bound and then shifted away by a numerical
artifact; it simply does not exist below the shelf of the canonical
self-consistent potential. The grid-converged direct eigenvalue count is
therefore
\begin{equation}
  N_{\mathrm{bound}}(\kap=1)=1.
  \nonumber
\end{equation}

The strongest correct statement is therefore that the canonical fully
screened self-consistent Poisson--Schr\"odinger accumulation-layer
problem has exactly one bound state, established numerically and
canonically by the self-consistent solution itself and supported both
by the sub-threshold WKB phase integral and by a direct grid-converged
eigenvalue count. This is not a general theorem for arbitrary
half-line potentials, but a sharply established property of the
canonical self-consistent solution family relevant to the present
problem.

The physical consequence is immediate. At \(\kap=1\), the accumulation
layer is a genuine one-level bound subsystem. The onset of
\(1-\kap>0\) therefore does not populate a second bound accumulation
state, but opens the extended channel. The fully screened endpoint is
thus already the clean limiting case of the later two-density picture:
one bound state carries the accumulation-layer sector, while all
additional charge belongs to the extended manifold. This is precisely
why the later orthogonality minimum, incoherent higher-subband sum, and
screening decomposition are so consequential: they reveal how the unique
bound state couples to an extended continuum-like sector under
incomplete screening.

The uniqueness statement at $\kap=1$ should be read together with the numerical stability of the stage-2 map discussed above. In the smooth regime relevant to the present application, the lifted problem converges to a reproducible canonical surface on the $(\kap,\lam)$ grid; outside that regime, discrete filling effects introduce visible kinks, but they do not invalidate the underlying canonical construction. The present appendix therefore establishes the scope of the stage-2 lift: a numerically stable and physically transparent canonicalization of the multisubband accumulation problem, together with a sharp spectral statement at its fully screened endpoint.

\bibliographystyle{apsrev4-2}
\bibliography{Literature_converted}

\end{document}